\documentclass{nature_figs}
\usepackage{amsmath}
\usepackage{amssymb}
\usepackage{hyperref}
\usepackage{lineno}
\usepackage{float}
\usepackage{graphicx}%
\usepackage{multirow}%
\usepackage{amsmath,amssymb,amsfonts}%
\usepackage{amsthm}%
\usepackage[title]{appendix}%
\usepackage{xcolor}%
\usepackage[super,comma]{natbib}
\usepackage{etoolbox}
\usepackage{textcomp}%
\usepackage{manyfoot}%
\usepackage{booktabs}%
\usepackage{xspace}
\usepackage{listings}%
\usepackage{adjustbox}
\usepackage{longtable}

\hypersetup{colorlinks,citecolor=blue,linkcolor=blue,urlcolor=blue}

\newcommand\psra{PSR~J0205$+$6449\xspace}
\newcommand\psrb{PSR~B2334$+$61\xspace}
\newcommand\velajr{CXOU~J0852$-$4617\xspace}

\newcommand\xmm{{\it XMM-Newton}\xspace}
\newcommand\chandra{{\it Chandra}\xspace}

\title{Constraints on the dense matter equation of state from young and cold isolated neutron stars}

\author{A.~Marino$^{*1,2}$, 
C.~Dehman$^{*1,2}$, 
K.~Kovlakas$^{*1,2}$, 
N.~Rea$^{*1,2}$, 
J.~A.~Pons$^{3}$, 
D.~Vigan\`o$^{1,2}$ 
}

\makeatletter
\let\saved@includegraphics\includegraphics
\AtBeginDocument{\let\includegraphics\saved@includegraphics}
\renewenvironment*{figure}{\@float{figure}}{\end@float}
\makeatother

\begin{document}

\maketitle

\begin{affiliations}

  \item Institute of Space Sciences (ICE), CSIC, Campus UAB, Carrer de Can Magrans s/n, E-08193, Barcelona, Spain
  \item Institut d'Estudis Espacials de Catalunya (IEEC), Edifici RDIT, Campus UPC, E-08860 Castelldefels (Barcelona), Spain
   \item Departament de F\'isica Aplicada, Universitat d’Alacant, Ap. Correus 99, E-03080 Alacant, Spain
 
\end{affiliations}


%
%

\begin{abstract}
Neutron stars are the dense and highly magnetic relics of supernova explosions of massive stars. The quest to constrain the Equation of State (EoS) of ultra-dense matter and thereby probe the behavior of matter inside neutron stars, is one of the core goals of modern physics and astrophysics. A promising method involves investigating the long-term cooling of neutron stars, comparing theoretical predictions with various sources at different ages. However, limited observational data, and uncertainties in source ages and distances, have hindered this approach. In this work, re-analyzing XMM-Newton and Chandra data from dozens of thermally emitting isolated neutron stars, we have identified three sources with unexpectedly cold surface temperatures for their young ages. To investigate these anomalies, we conducted magneto-thermal simulations across diverse mass and magnetic fields, considering three different EoS. We found that the "minimal" cooling model, failed to explain the observations, regardless the mass and the magnetic field, as validated by a machine learning classification method. The existence of these young cold neutron stars suggests that any dense matter EoS must be compatible with a fast cooling process at least in certain mass ranges, eliminating a significant portion of current EoS options according to recent meta-modelling analysis.
\end{abstract}


Neutron stars are incredibly dense objects with densities several times that of atomic nuclei ($\rho\sim10^{14}$\,g~cm$^{3}$). They hold unique information about the properties and behaviour of matter under extreme conditions of densities and magnetic fields \cite{Baym1971,Pacini1967,HardingLai2006}. Their internal structure, mass-radius relationship, and overall behaviour relies on a unique equation of state (EoS), which describes the relationship between pressure, density, and composition, in a regime unreachable on Earth laboratories. The EoS not only determines the structure, cooling rates, and rotational properties of neutron stars, but it also plays a role in astrophysical phenomena such as gravitational wave signals emitted during mergers with other neutron stars or black holes. Deciphering what is the actual EoS of dense matter is a key open question for several branches of physics.\\
Constraining the dense matter EoS involves the combination of various theoretical models, computational techniques, and astrophysical observations, all aimed at refining and validating our understanding through the comparison of theoretical predictions with observational data. The interactions between particles, such as neutrons, protons, and electrons, are crucial factors that shape the EoS at different density regimes, as are the superfluid components. Furthermore, as densities increase toward the core of the star, the nature of matter within neutron stars becomes more uncertain, with the possible appearance of exotic particles like hyperons, meson condensates, or quark matter \cite{Prakash1997}. 

Neutron star cooling is caused by a combination of neutrino emission from the dense core of the star and thermal photons emitted from the outer layers. By measuring the surface temperatures of many objects over a large age range, neutron star cooling models (and consequently, the EoS) can be constrained \cite{Page2004}. The cooling history of a neutron star involves multiple parameters a part from the EoS, leading to diverse tracks for neutron stars born with different initial conditions, such as the object mass, the envelope composition and the initial magnetic field. 

Cooling curves were historically divided into two theoretical classes: 1) the "standard" or "minimal" cooling ones, dominated by modified Urca processes possibly with the addition of superconductivity or superfluidity in the core producing neutrino cooling via Cooper pairs \cite{Flowers1976}; 2) those showing "enhanced cooling" due to the activation of direct Urca processes, hyperons, or even quark matter or meson condensates \cite{Prakash1992,Prakash1998}. \\
Evidences for the presence of enhanced cooling were presented for some isolated neutron stars\cite{Page2004, Ho2009, Beznogov2015, potekhin2015, Potekhin2018}, and for transiently accreting neutron stars\cite{Jonker2007, Heinke2007, Brown2018, Mendes2022, Potekhin2023}. Both scenarios can provide independent constraints on the neutron star cooling. However, on one side the lack of well-constrained spectral energy distributions, exact ages and/or precise distances, and on the other side uncertainties on the accretion state and history, were precluding thus far any firm and conclusive constraint on the EoS. Furthermore, it was suggested \cite{Aguilera2008} that the dissipation of the magnetic field in the highly resistive crust could hide the effect of the enhanced cooling mechanisms for stars with magnetic fields above $10^{14}$\,G. 
In this work, we present a detailed study of three extremely cold, young, and close-by neutron stars, the mere existence of which is constraining the neutron star EoS because the “enhanced” cooling processes are required to reconcile the models with the observational data. 

\begin{figure*}
\centering
\includegraphics[width=0.95\textwidth]{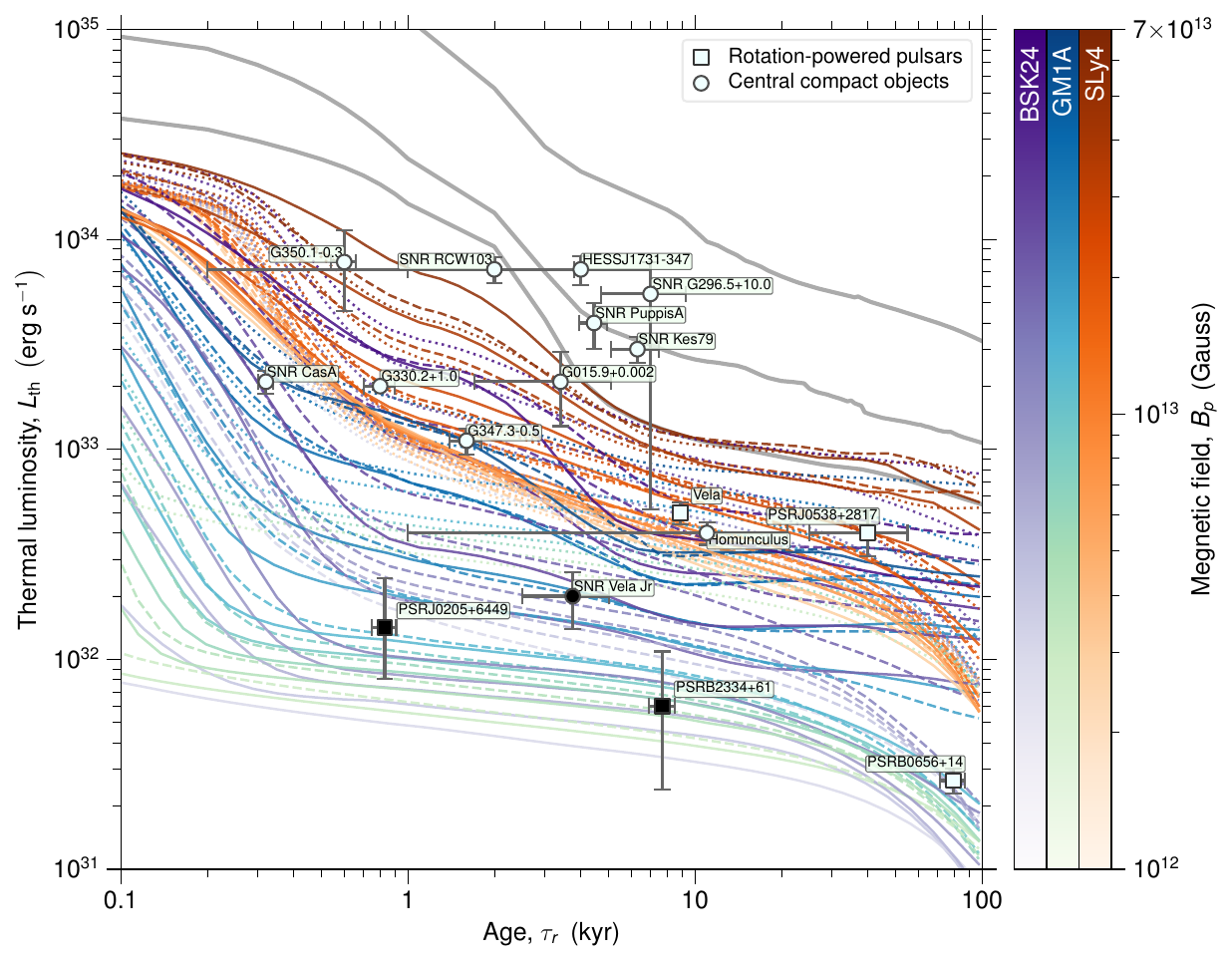}
\caption{Comparison between observational data and theoretical cooling curves. 
Observations include rotational powered pulsars (squares), central compact objects (circles), with the three sources studied in this work in black. The 81 cooling curves used in our analysis are coloured based on their EoS: SLy4 (orange), BSK24 (violet), GM1A (blue). The line styles denote the masses: 1.4\,M$_\odot$ (dots), 1.6\,M$_\odot$ (dashed lines), and 1.8\,M$_\odot$ (solid lines). The colorbars show the dipolar magnetic field values which were in the range of $B_p = 1-70\times10^{12}$\,G. Only for comparison, we also show three grey curves corresponding to stronger magnetic field intensities of $10^{14}$\,G, $3\times 10^{14}$\,G, and $10^{15}$\,G, not used in our statistical analysis. 
}
\label{fig:cooling}
\end{figure*}

To perform a systematic study of surface temperatures in isolated neutron stars across different classes, we re-analysed deep \xmm and/or \chandra data for a sample of 70 isolated neutron stars. Such a sample consists of sources with a statistically significant spectral contribution by a thermal component and with a reliable and sufficiently well-known age and distance (Dehman et al. 2024, in prep.). 
Out of this sample, three sources stood out for being colder by about an order of magnitude with respect to the other objects at similar young ages (see Extended Data Table\,\ref{tab:par}, \ref{tab:fit_results}, and \ref{tab:log}). These sources are two rotation-powered pulsars (RPPs) \psra ($P=70$\,ms, $B_p = 7\times10^{12}$\,G and Age$=841$\,yr\cite{Kothes2013}) and \psrb\, ($P=490$\,ms, $B_p = 2\times 10^{13}$\,G and Age$=7700$\,yr\cite{YarUyaniker2004}), and a central compact object (CCO), \velajr (Age=$2500-5000$ yr\cite{Allen2015}). All three objects have an associated supernova remnant studied at all wavelengths providing precise distance and age constraints (in the case of \psra also the historical record of its associated supernova, SN 1118; see also Methods). For each source, we estimated the effective black body temperature and the radius of the emitting surface through X-ray spectral analysis and used these values to calculate their thermal luminosity. We used simple black body models (see Extended Data Figure\,\ref{fig:spec} and Extended Data Table\,\ref{tab:fit_results}), but checked that our final results held using more sophisticated atmosphere models (Extended Data Table\,\ref{tab:fit_atmo_results}) and considering the presence of additional "hidden" thermal contribution by the whole NS surface (see Methods, Extended Data Figure\,\ref{fig:coldbb} and Extended Data Figure\,\ref{fig:luminosity}). 
In Figure~\ref{fig:cooling} we report on the thermal luminosities of all the RPPs and CCOs having a precise estimate of the age and distance, the former not relying on the characteristic age of the pulsar inferred from the timing analysis (which is known to have large uncertainties). Being so young, i.e., with ages between $800 - 8000$~yrs, and yet so cold, \psra, \psrb and \velajr must have necessarily experienced some sort of rapid cooling. \\
To comprehend why the thermal X-ray luminosity appears significantly faint during the early stages of these three objects, we conducted magneto-thermal simulations\cite{pons2009,vigano2013,vigano2021} exploring different EoSs, masses, and a range of magnetic-fields.
In particular, our collection of 81 simulations comprises three distinct EoSs with various cooling channels, including modified Urca, Cooper pairs and direct Urca (among other channels; see e.g. \cite{potekhin2015}). In particular, SLy4 \cite{douchin2001} assumes a standard ``minimal'' cooling scenario and does not activate “enhanced cooling” processes, while BSK24 \cite{pearson2018} and GM1A\cite{gusakov2014,anzuini2021,anzuini2022}, for certain masses, do involve “enhanced cooling” processes such as nucleon direct Urca and hyperons direct Urca (the latter activated only for the GM1A).

Additionally, we considered three different masses (1.4, 1.6, and 1.8 M$_\odot$), along with nine initial magnetic field values for the surface dipolar field ranging from $1\times10^{12}$\,G to $7\times10^{13}$\,G at the pole, with no toroidal magnetic field to avoid Joule heating and focus on the cold neutron star scenario. Furthermore, we solely employed an iron envelope model, as alternative compositions containing light elements would predict at these ages a thermal luminosity approximately one order of magnitude brighter than that projected by the iron-envelope model \cite{dehman2023b} (see the Methods and Extended Data Figure\,\ref{fig:envelope-comp}).

\begin{figure*}
\centering
\includegraphics[width=0.53\textwidth]{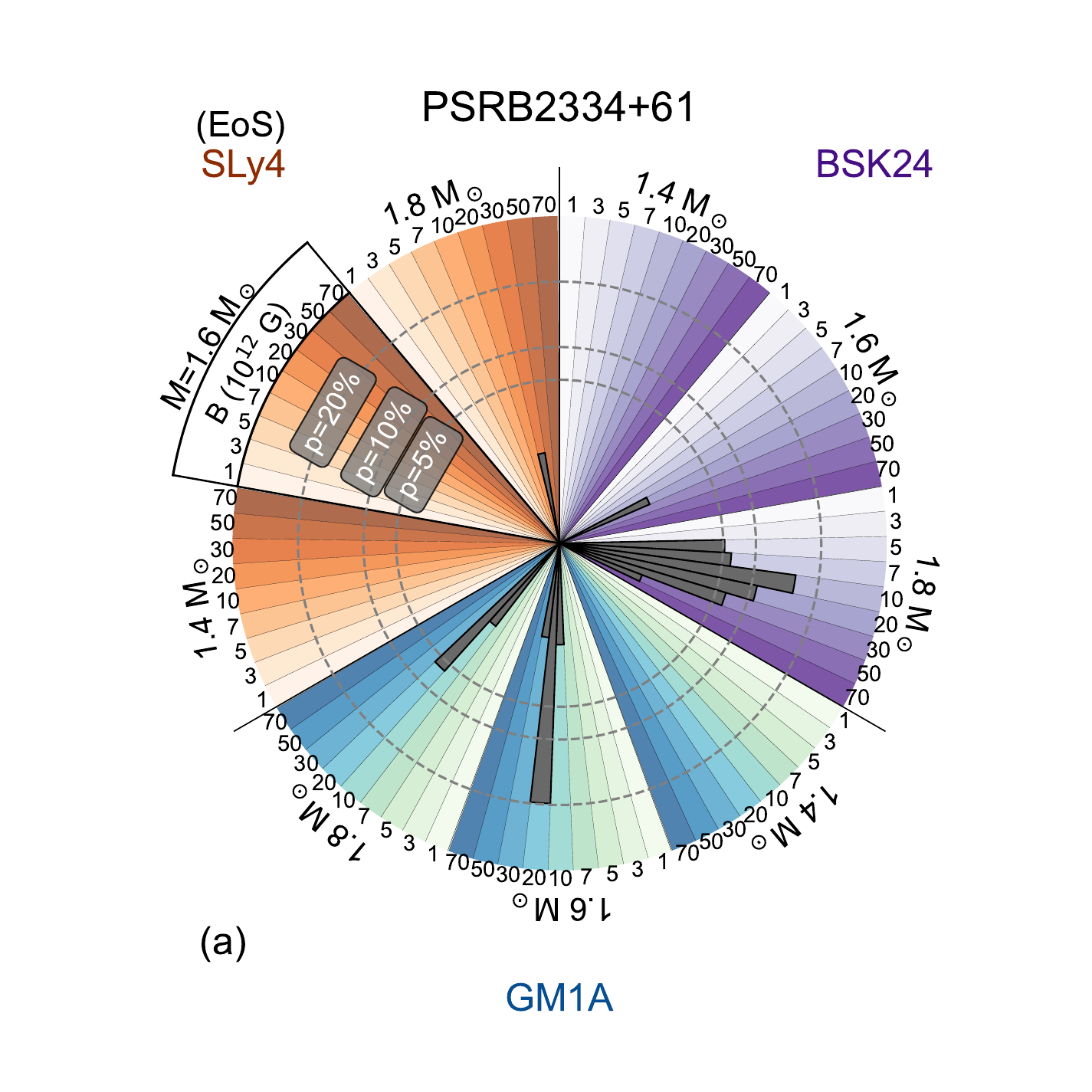}
\hspace{-1.5cm}
\includegraphics[width=0.53\textwidth]{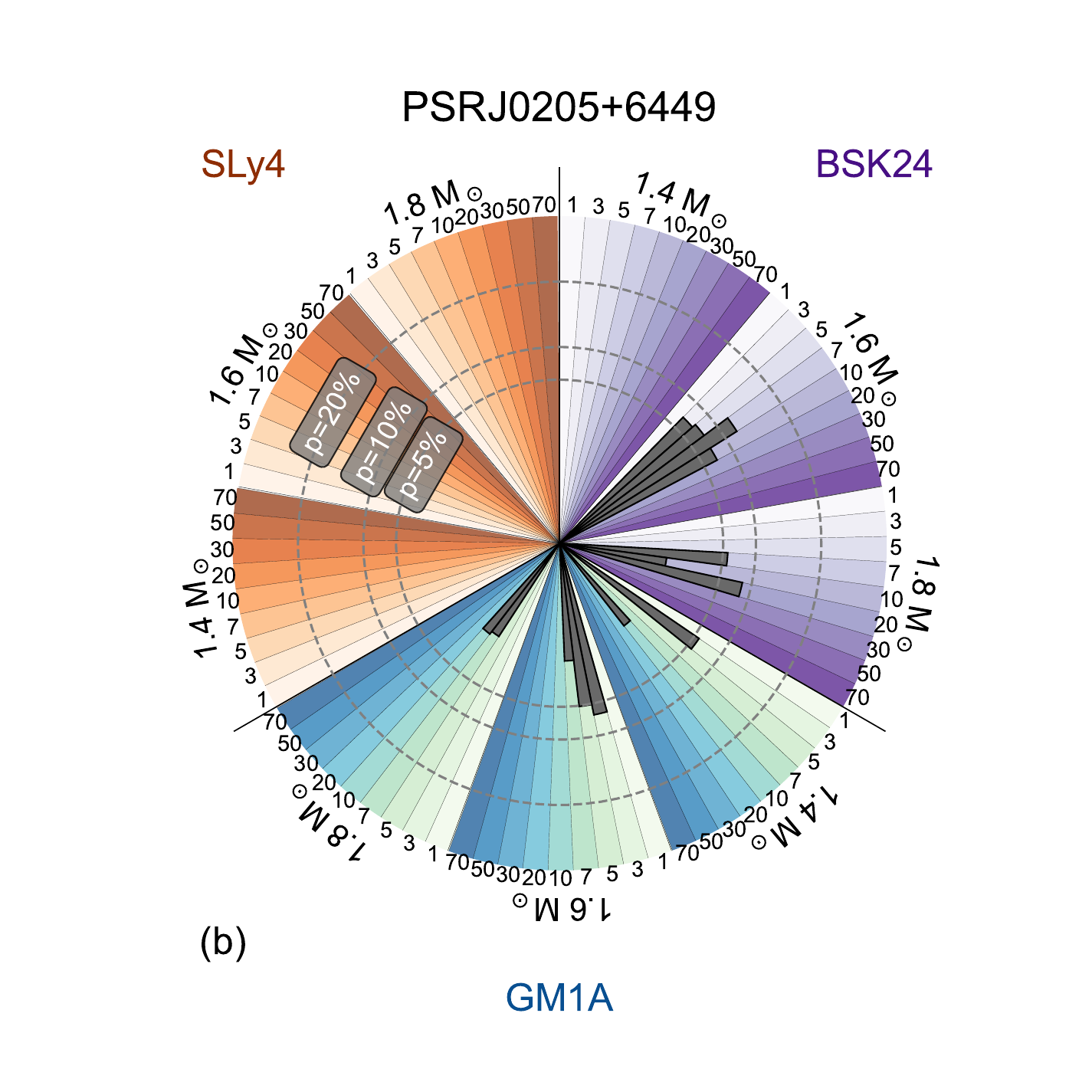}
\caption{Chart pies breaking down the results of our 4D classification method for \psrb (left) and \psra (right). Each chart is divided in three sectors for each of the EoS used: BSK24 (purple), SLy4 (orange), GM1A (green). Each sector is further broken down by mass and magnetic field. Superimposed to the cloves, grey histograms denote the posterior probability for each simulation-clove to be the cooling curve followed by the source.}\label{fig:statistical}
\end{figure*}

In Figure~\ref{fig:cooling} we show the observational measurements compared with the magneto-thermal evolutionary tracks. Even at first glance, it is clear that some of the explored scenarios are not matching the faint thermal luminosities of these extremely cold sources. Indeed, assuming a SLy4 EoS, the dramatic drop in surface temperature in the three outliers could not be reached for any combination of mass and magnetic fields. In the exotic case of having hyperons in the core, namely the case of GM1A, the cooling might proceed fast enough to be compatible with the observational data. On the other hand, for the BSK24 EoS, when the mass is larger than $1.6 M_{\odot}$ direct Urca is activated and the tracks show an enhanced cooling compatible with the data. Following a less qualitative and more rigorous approach, we used machine learning (see Methods) to find the range of parameters that better described each source. At that aim, we first considered the observational data and the simulations in a 3D space considering as independent parameters the thermal luminosity $L_{\rm th}$, spin period $P$ and spin derivative $\dot{P}$. We then also extended these simulations into a 4D space, including also the age of the sources. Since for the CCO \velajr , $P$ and $\dot{P}$ are unknown, this analysis was carried out only for the two RPPs. The extension to a 4D space allowed us to check whether curves that may explain the observed $L_{\rm th}$ would also predict $P$ and/or $\dot{P}$ compatible with the timing parameters of the sources at the same age. \\
In Figure~\ref{fig:statistical} we summarise in a chart pie the results of our machine learning simulations in 4D (the results of the 3D are very similar; see Methods, Extended Data Table\,\ref{tab:classprob4D} and Extended Data Figure\,\ref{fig:3dplots}). 
According to these methods, we can quantitatively exclude the EoS without a physical mechanism to activate an enhanced cooling at young age (in our set of EoSs this would correspond to SLy4). \\
In particular, for \psrb all cooling curves point toward the source having a relatively high mass ($\sim1.6M_{\odot}$) and a dipolar magnetic field at birth of $\sim0.7-3\times10^{13}$\,G, compatible with its current value. Despite a slight preference is visible for the hyperon EoS (GM1A) for this pulsar, we caveat that more simulations with other enhanced cooling EoSs might also reach similar probabilities, hence at this stage, we cannot constrain the exact EoS with this technique. 
Applying these methods to \psra, we find that both "enhanced cooling" EoSs, BSK24 and GM1A, with masses of $\sim1.6M_{\odot}$ are compatible with its observed parameters, resulting in an initial magnetic field within $\sim3-20\times10^{12}$\,G, again in agreement with the estimated value for this pulsar. For both sources, the simulations with the SLy4 EoS considering only "minimal" cooling for any set of parameters provide a non-acceptable match with the data (probabilities $<5\%$; see also Methods). Furthermore, considering only the source thermal luminosity and age, this conclusion holds also for \velajr. 

Given the uniqueness of the EoS, these results provide evidence that neutron stars cannot be governed by an EoS that is not compatible with the low luminosities of \psra, \psrb, and \velajr\, at least in a certain mass range. The precise error determinations, the updated values for the distances and/or the ages of these three sources, along with the machine learning approach employed to corroborate our conclusions, make these results the strongest measurement up to now in favour of enhanced cooling.  We found that only EoSs (and compositions) allowing a fast-cooling process in the first few kyrs can be successfully reconciled with the thermal emission of all sources in our sample.
While a detailed and comprehensive analysis of various possibilities for fast cooling (e.g., hyperons, quarks, pure nucleonic matter with very large symmetry energy) is beyond the scope of this paper, the issue itself stimulates thought. In particular, considering a simplified nucleonic meta-modelling\cite{margueron2017} the proposed EoS that does not have a high enough proton fraction to activate fast cooling processes for any reasonable neutron star mass, is estimated to be about 75\%. This modelling, despite being simplified and dependent on the assumed composition, shows that a significant fraction of the currently available EoS are potentially excluded by the mere existence of these cold and young neutron stars.

\begin{addendum}
 \item AM, CD and NR are supported by the ERC Consolidator Grant 'MAGNESIA' No. 817661 (PI: N. Rea) and SGR2021.01269 funded by the Generalitat de Catalunya (PI: V. Graber/N. Rea). CD acknowledges the doctoral program in Physics of the Universitat Aut\`onoma de Barcelona and the Nordita Visiting PhD Fellow program. KK is supported by a fellowship program at the Institute of Space Sciences (ICE-CSIC) funded by the program Unidad de Excelencia Mar\'ia de Maeztu CEX2020-001058-M. JAP acknowledges support from the Generalitat Valenciana grants ASFAE/2022/026 (with funding from NextGenerationEU PRTR-C17.I1) and CIPROM/2022/13, and from the AEI grant PID2021-127495NB-I00 funded by MCIN/AEI/10.13039/501100011033 and by “ESF Investing in your future”. DV is supported by the ERC Starting Grant 'IMAGINE' No. 948582 (PI: D. Vigan\`o). 

 KK thanks the ``Summer School for Astrostatistics in Crete'' for providing training on the statistical methods adopted in this work.
 We thank F. Anzuini for providing his Hyperon EoS implementation and for detailed comments on the manuscript, as well as L. Tolos, C. Manuel, D. De Grandis, S. Ascenzi, V. Graber, F. Gulminelli, M. Oertel, A. Borghese, and F. Coti Zelati for feedback and comments.

  \item[Author Contributions]The project was conceived, planned and organized by NR, who wrote the final manuscript. The three main parts of the work have been performed by AM, CD and KK, namely the observational data analysis, the magneto-thermal simulations, and the machine learning statistical analysis, respectively. They were also in charge of writing the relative sections in the Methods. JAP, DV, and CD wrote the magneto-thermal simulation code employed in this work. All authors discussed and commented on the final complete manuscript.
  
  \item[Competing Interests] The authors declare that they have no competing financial interests.
 \item[Data Availability]  Data that supports this paper are publicly available in the {\em XMM-Newton} and {\em Chandra} archives. Further data products can be supplied by the authors on request.
  \item[Code Availability] Codes that support this paper are available upon request to the authors.
   \item[Correspondence] Correspondence and requests for materials should be addressed to the four corresponding authors: A.M., C.D., K.K, and N.R (marino@ice.csic.es, dehman@ice.csic.es, kovlakas@ice.csic.es, rea@ice.csic.es)

\end{addendum}

\newpage

\newcommand{\actaa}{Acta Astron.}   
\newcommand{\araa}{Annu. Rev. Astron. Astrophys.}   
\newcommand{\areps}{Annu. Rev. Earth Planet. Sci.} 
\newcommand{\aar}{Astron. Astrophys. Rev.} 
\newcommand{\ab}{Astrobiology}    
\newcommand{\aj}{Astron. J.}   
\newcommand{\ac}{Astron. Comput.} 
\newcommand{\apart}{Astropart. Phys.} 
\newcommand{\apj}{Astrophys. J.}   
\newcommand{\apjl}{Astrophys. J. Lett.}   
\newcommand{\apjs}{Astrophys. J. Suppl. Ser.}   
\newcommand{\ao}{Appl. Opt.}   
\newcommand{\apss}{Astrophys. Space Sci.}   
\newcommand{\aap}{Astron. Astrophys.}   
\newcommand{\aapr}{Astron. Astrophys. Rev.}   
\newcommand{\aaps}{Astron. Astrophys. Suppl.}   
\newcommand{\baas}{Bull. Am. Astron. Soc.}   
\newcommand{\caa}{Chin. Astron. Astrophys.}   
\newcommand{\cjaa}{Chin. J. Astron. Astrophys.}   
\newcommand{\cqg}{Class. Quantum Gravity}    
\newcommand{\epsl}{Earth Planet. Sci. Lett.}    
\newcommand{\frass}{Front. Astron. Space Sci.}    
\newcommand{\gal}{Galaxies}    
\newcommand{\gca}{Geochim. Cosmochim. Acta}   
\newcommand{\grl}{Geophys. Res. Lett.}   
\newcommand{\icarus}{Icarus}   
\newcommand{\ija}{Int. J. Astrobiol.} 
\newcommand{\jatis}{J. Astron. Telesc. Instrum. Syst.}  
\newcommand{\jcap}{J. Cosmol. Astropart. Phys.}   
\newcommand{\jgr}{J. Geophys. Res.}   
\newcommand{\jgrp}{J. Geophys. Res.: Planets}    
\newcommand{\jqsrt}{J. Quant. Spectrosc. Radiat. Transf.} 
\newcommand{\lrca}{Living Rev. Comput. Astrophys.}    
\newcommand{\lrr}{Living Rev. Relativ.}    
\newcommand{\lrsp}{Living Rev. Sol. Phys.}    
\newcommand{\memsai}{Mem. Soc. Astron. Italiana}   
\newcommand{\maps}{Meteorit. Planet. Sci.} 
\newcommand{\mnras}{Mon. Not. R. Astron. Soc.}   
\newcommand{\nat}{Nature} 
\newcommand{\nastro}{Nat. Astron.} 
\newcommand{\ncomms}{Nat. Commun.} 
\newcommand{\ngeo}{Nat. Geosci.} 
\newcommand{\nphys}{Nat. Phys.} 
\newcommand{\na}{New Astron.}   
\newcommand{\nar}{New Astron. Rev.}   
\newcommand{\physrep}{Phys. Rep.}   
\newcommand{\pra}{Phys. Rev. A}   
\newcommand{\prb}{Phys. Rev. B}   
\newcommand{\prc}{Phys. Rev. C}   
\newcommand{\prd}{Phys. Rev. D}   
\newcommand{\pre}{Phys. Rev. E}   
\newcommand{\prl}{Phys. Rev. Lett.}   
\newcommand{\psj}{Planet. Sci. J.}   
\newcommand{\planss}{Planet. Space Sci.}   
\newcommand{\pnas}{Proc. Natl Acad. Sci. USA}   
\newcommand{\rpp}{Rep. Prog. Phys.}
\newcommand{\procspie}{Proc. SPIE}   
\newcommand{\pasa}{Publ. Astron. Soc. Aust.}   
\newcommand{\pasj}{Publ. Astron. Soc. Jpn}   
\newcommand{\pasp}{Publ. Astron. Soc. Pac.}   
\newcommand{\raa}{Res. Astron. Astrophys.} 
\newcommand{\rmxaa}{Rev. Mexicana Astron. Astrofis.}   
\newcommand{\sci}{Science} 
\newcommand{\sciadv}{Sci. Adv.} 
\newcommand{\solphys}{Sol. Phys.}   
\newcommand{\sovast}{Soviet Astron.}   
\newcommand{\ssr}{Space Sci. Rev.}   
\newcommand{\uni}{Universe} 

\newpage


\begin{methods}

\subsection{Source sample.}\label{sec:sample}
In this work we present the analysis of deep X-ray observations of three neutron stars, \psra, \psrb, and \velajr, as part of a larger work done on the re-analysis of a total of 70 isolated thermally emitting neutron stars (Dehman et al. 2024 in prep). The former two sources are young radio pulsars, whose emission is directly powered by their rotational energy, so that they are classified as RPPs. \velajr is instead a non-pulsating neutron star classified as a CCO, a class of isolated neutron stars localised in the the geometrical centre of their SNR. A summary of their timing parameters, the available X-ray observations and all relevant references are presented in the Extended Data Table \ref{tab:log} and \ref{tab:par}. 

All neutron stars in our sample are associated to a SNR. \psra is associated to the pulsar wind nebula 3C 58 and to the SNR G130.7+3.1, and it is emitting from radio to gamma-rays \cite{Kuiper2010, Abdo2009}. The other pulsar in the sample, \psrb , is also associated to a SNR, i.e. to SNR G114.3+0.3 \cite{Furst1993}, and its detected in radio and X-rays.
The CCO \velajr is the X-ray bright point-source lying only $\sim$35 arcseconds away \cite{Camilloni2023} from the geometrical center of the shell-type SNR G266.2-1.2, also known as Vela Jr\cite{Aschenbach1998}. As many other CCOs it does not have detected pulsations, but its faint optical magnitude limits, lack of variability and X-ray spectral properties confirm its neutron star nature\cite{Kargaltsev2002}. Interestingly, \velajr is one of the very few CCOs that might have been observed in a different waveband than the X-rays. A candidate faint point-like infrared counterpart for the source has been found with ESO-VLT observations\cite{Mignani2019,Camilloni2023} and interpreted as emission from the neutron star magnetosphere or from a relic disc around the CCO .
None of the three sources showed any variability in our data (nor in the past), as it is indeed expected for an isolated neutron star with a relatively low magnetic field.


\subsection{Distances and real age constraints}\label{sec:dist-age}

The location on the luminosity-age plane (see Figure~\ref{fig:cooling}) is critically dependent on the accuracy with which we know both the distance and the age of our sources. In the three sources we report here, the SNRs and HI regions around the objects were studied in great detail, providing robust measurements of distances and ages.

\psra is considered one of the youngest pulsars known. The source has been proposed to be the leftover of the historical supernova SN 1181, providing an age of 839 years. However, several elements, such as the measured expansion speeds of both the synchrotron bubble and of the thermal filaments \cite{Bietenholz2006}, initially suggested that \psra (and 3C 58) may be instead older by thousands of years, than SN 1181. Note that even in the unlikely case that \psra is older than SN 1181, an upper limit on its age can be posed using its characteristic age $\tau_c$ of 54 kyr. Even assuming this upper limit, the SLy4 curves cannot explain its luminosity. Its distance was originally estimated to be 3.2 kpc \cite{Roberts1993}. More recent HI measurements have instead placed the source at a closer distance of just 2\,kpc \cite{Kothes2013, Ranasinghe2022}. This is in-line with the association with SN1181, and compatible with the source proper motion \cite{Bietenholz2013}. It is noteworthy, however, that the results of this manuscript would hold even assuming the outdated distance of 3.2 kpc, as the corresponding increase in luminosity would still place \psra at a value of$\lesssim$10$^{33}$ erg/s, i.e., lower than the cooling curves simulated with SLy4.


For \psrb instead, a distance range of 2.1--3.3\,kpc was reported by several authors \cite{Cordes2002,McGowan2006} according to its radio Dispersion Measure (DM). In the following we adopt those values. It is noteworthy that this result has been recently challenged by observations of the HI line in the SNR G114.3+0.3, according to which the source (and its SNR) is placed at a much closer location, i.e. at 700\,pc \cite{YarUyaniker2004}. Additionally, a 0.1-0.9 kpc range was reported by \cite{Ranasinghe2022} on the basis of kinematic analysis. Note that if this distance is assumed the pulsar would be even colder, hence our conclusions will not change.
A value of 7.7 kyr was estimated for the real age of the source from the study of the SNR  \cite{YarUyaniker2004} (while its characteristic age is $\tau_c$ is 40 kyr).
The real age of \psrb has been reported without any uncertainty. We therefore adopted a 10\% relative error on this value throughout this manuscript. For consistency, we also used a 10\% uncertainty for the age of \psra, which is lacking of uncertainties as well since it coincides with the historical date of SN 1181.

Finally, \velajr is associated to the SNR Vela Jr (for a critical discussion on the topic, see \cite{Kargaltsev2002}). Several arguments, including its expansion rate, the estimated shock speed, its association to the Vela Molecular Ridge (see, for a review \cite{Allen2015}), suggest that \velajr distance lies in the 0.5-1.0 kpc range, and its real age would lie in the range 2.5-5.0 kyr \cite{Allen2015}.


\subsection{Data reduction}
\label{sec:data-red}

In this work, we only considered \xmm\ and \chandra\ observations\cite{Slane2004,McGowan2006,Kargaltsev2002}, being those providing the most accurate X-ray spectral energy distributions. Details on all observations can be found in Extended Data Table \ref{tab:log}. \\
\psra and its nebula 3C 58 have been observed three times by \chandra between 2001 and 2003 \cite{Slane2004}. \psrb has been observed only once by \xmm \cite{McGowan2006}. On the other hand, many archival \xmm and \chandra observations are available for \velajr \cite{Kargaltsev2002}. For the latter, we only analysed the \chandra observation with the longest exposure time, and for which the Advanced CCD Imaging Spectrometer ACIS-S has been employed. We used only the \chandra\ data to excise more effectively any contribution from the underlying SNR.  \\
In the following subsections, we will describe the data reduction procedure followed for all the data sets used in this work.

\subsection{XMM-Newton}
We include data from the EPIC-pn detector \cite{Struder2001} onboard \xmm. For all the observations, the EPIC-pn was set in Small Window mode (time resolution of 5.7\,ms). Data reprocessing was performed with the \xmm Science Analysis Software (SAS) v. 20.0.0. In order to reduce the data, we first filtered the event files for periods of high background activity. We used the \textsc{epatplot} tool to display the observed pattern distribution versus the expected one and thereby assess the pile-up impact, finding that it was negligible for all the observations considered here. The source counts were extracted from a circle of 20 arcsec radius centred on the coordinates of the source. For the background, we used a region of the same size and shape, located sufficiently far from the source.

\subsection{Chandra}
All three \chandra observations of \psra and the single observation of \velajr used here were carried out using the ACIS-S. Data processing, reduction and spectral extraction of all observations was performed within \textsc{CIAO} v.4.14 using the standard pipelines. A 3x3 pixel square box was used to identify source photons, small enough to limit contamination from the surrounding nebula, following standard approaches\cite{Slane2004}. A region of the same size, but located far from the source and 3C 58 was used for the background. The routine \textsc{specextract} task was used to build the spectra and create the ancillary response and the redistribution matrix files.

\subsection{Spectral analysis}\label{sec:spectral}
The spectral analysis was performed using \textsc{Xspec} v.12.12.1. For \psra and \velajr, the spectra were grouped in order to have at least 20 counts per bin, enabling use of the $\chi^2$ statistics. The low count-rate of the \psrb observation forced us to choose a different grouping strategy, 10 counts per bin, and consider the C-statistics instead. We kept data only within an energy interval where the source data points were higher than the background level, thereby selecting 0.3-10 keV, 0.5-2 keV and 0.5-4.5 keV for \psra , \psrb and \velajr, respectively. In all the models used to describe the data we included the \textsc{tbabs} component to take into account the effect of  the interstellar absorption, setting the photoelectric cross-sections to the values provided by \cite{Verner1996}. For the elemental abundances, we tried both the tables by \cite{Wilms2000} and by \cite{Anders1989} finding negligible differences.

While the spectral shapes of \psrb and \velajr were relatively simple, being consistent with a single, thermal component, the spectrum of \psra is apparently dominated by a power-law component. Details on the results of the fits are reported in Extended Data Table \ref{tab:fit_results}, while we refer to Extended Data Figure\,\ref{fig:spec} for a plot with spectra, best-fit models and corresponding residuals. 

For \psra, we initially used a simple \textsc{powerlaw} model to describe the three available \chandra\ spectra, finding an already acceptable fit ($\chi^2$/d.o.f.=587/600). We then added a blackbody component, described with the \textsc{bbodyrad} model in \textsc{Xspec}. The parameters of \textsc{bbodyrad} are the blackbody temperature $kT_{\rm bb}$ and its normalisation $K_{\rm bb}$, which translates into the blackbody radius $R_{\rm bb}$ by means of the formula $K_{\rm bb}=\left(R_{\rm bb}/D_{\rm 10 \ kpc}\right)^2$, with $D_{\rm 10 \ kpc}$ the distance of the source in units of 10~kpc. Adding a blackbody component, results in a slight improvement in the fit, which turns out to be highly significant when we use the \cite{Anders1989} abundances for $N_{\rm H}$ whereas only marginally significant with the \cite{Wilms2000} abundances, with probability of improvement by chance of $\sim10^{-7}$ and $\sim10^{-3}$, estimated using \textsc{ftest}. Such a result indicates the clear presence of correlation between $N_{\rm H}$ and the blackbody parameters. In order to further investigate the significance of the thermal component in the fit with the \cite{Wilms2000} abundances we applied the Goodman-Weare algorithm of Markov Chain Monte Carlo (MCMC; \cite{Goodman2010}) to produce contour plots for $N_{\rm H}$, $kT_{\rm bb}$ and $R_{\rm bb}$. We used 20 walkers and a chain length of 5$\times$10$^5$, to calculate the marginal posterior distributions of the best-fit parameters for all three \chandra spectra. The results are presented in the corner plot in Supplementary Data Figure 1.
As evident, both $kT_{rm bb}$ and $K_{\rm bb}$ can be sufficiently constrained and the normalisation is small, but still not consistent with zero. Such a result confirms the presence of a thermal component in the emission of \psra , in agreement with the results from previous studies of the source where the thermal emission was indeed significantly detected\cite{Slane2004}. 

In order to describe the emission by \psrb , we tried both a \textsc{bbodyrad} model or a single \textsc{powerlaw}. Despite providing a statistically acceptable fit, the latter model requires a very soft power-law component, with $\Gamma > 6$, which is unphysical and suggests that the power-law component is actually mimicking a blackbody-like spectrum. We therefore us a simple \textsc{bbodyrad} model (similar to \cite{McGowan2006}) which is also sufficient to describe the spectrum of the thirs source, \velajr, as also reported by other authors\cite{Kargaltsev2002}. The best-fit models and relative residuals for these sources are shown in Extended Data Table \ref{tab:fit_results} and Extended Data Figure\,\ref{fig:spec}.
According to these fits, the thermal emissions from \psra and \psrb are both consistent with blackbody temperatures of 0.15-0.25\,keV (with the former slightly hotter) arising from hot spots of size 1-2 km. Interestingly, \velajr is instead characterised by a significantly hotter blackbody temperature, i.e. around 0.4 keV, and smaller size, about 0.2-0.3 km. The results of the fits are consistent, within the errors, with studies using the same data sets and absorbed blackbody models\cite{Slane2004,McGowan2006,Kargaltsev2002}.

Once the best-fit models were found, we applied the convolution model \textsc{cflux} to the \textsc{bbodyrad} component to estimate the bolometric flux that can be considered purely thermal. These results are reported in Extended Data Table \ref{tab:fit_results}. 

We then used the source distances to estimate their thermal luminosity following standard propagation errors techniques. In the following will be referred to as observed luminosity $L_{\rm obs}$. However, for our statistical analysis, we estimated the error distribution for \psra and \psrb using a more rigorous approach that involves Monte Carlo uncertainty propagation (see below). The luminosity obtained with this technique, $L_{\rm stat}$ from now on, is always well compatible within the errors with $L_{\rm obs}$. Both values are reported in Supplementary Data Table 1.

\subsection{Testing atmosphere models}
In this work, we have used systematically a simple blackbody model to describe the thermal emission from these sources. Despite its success in describing the spectra from several isolated neutron stars and its simplicity, this approach ignores some crucial details on the underlying physics of NS emission: the possible presence of an athmosphere which modifies the blackbody emission coming from the NS surface through Compton scattering and other radiative processes. Several studies have shown that when the thermal component is modelled using a proper model for the NS atmosphere instead of a blackbody model, as in the case of Cas A\cite{Ho2009,Heinke2010}, a different set of parameters could be obtained, although at the expenses of assuming certain densities and compositions, which are typically unknown. In order to check whether our results would change when adopting atmosphere models, we have fitted our three sources with \textsc{nsmaxg} \cite{Ho2009}, which is suited to probe different chemical compositions (H, C, O) and different magnetic fields intensities. We first tried to fix the normalization $N_{\rm nsmaxg}$, which is connected to the radius of the emitting region with the formula $N_{\rm nsmaxg}=(R_{\rm em}/R_{\rm NS})^2$, to 1, corresponding to the case where the whole surface emits. However, the resulting fits were unacceptable, so we left $N_{\rm nsmaxg}$ free to vary, finding always values lower than 1. We also fixed the distances to the values reported in Extended Data Table \ref{tab:par}. For all sources we explored the case of $B=10^{12} G$ and tried as chemical compositions H, C and O. For the two RPPs, we found that all the probed chemical compositions could lead to equally acceptable fits with respect to the simple blackbody models. For \velajr , however, only the model with H was compatible with the data. Finally, once the best-fit parameters were obtained, we used again \textsc{cflux} to estimate the bolometric flux. We report the final parameters in Extended Data Table \ref{tab:fit_atmo_results}. From a comparison between the results with a simple blackbody model (Extended Data Table \ref{tab:fit_results}), we find that indeed adopting more physically motivated models of the NS surface emission leads to different best-fit values for the effective temperatures and the sizes of the emitting regions. Nevertheless, using an atmosphere model leads in all cases to bolometric fluxes which are systematically smaller or at least compatible within the errors with respect to the fluxes estimated through a simple blackbody model. These tests show that the low thermal luminosities reported in this manuscript might even represent an overestimate of the actual thermal luminosity from these sources, solidifying our conclusions.

\subsection{Adding the hidden contribution from the whole NS surface} The thermal components detected in all these sources are characterised by blackbody radii of a few km or less and can be therefore considered as being radiated by hot spots on the NS surface. However, a contribution from the entire NS surface at lower temperature, i.e., in an energy range where interstellar absorption is stronger, can not be excluded a priori. We therefore tried to estimate the upper limit of this hidden contribution by adding to our final fits an additional blackbody component with radius fixed to the NS assumed radius (in our case a 11 km\cite{Ozel2016} value was chosen). The temperature of this blackbody component was used to estimate an upper limit on the thermal luminosity from the entire surface, $L_{\rm surf}$ hereafter. 

We show both the best-fit and the upper limit of a possible undetected surface blackbody thermal component in Extended Data Figure~\ref{fig:coldbb}, displaying how even if actually present, the latter emission is always made undetectable by interstellar extinction.
For all sources, these upper limits, reported in Supplementary Data Table 1, are still too low to reconcile these sources with the cooling tracks simulated with the SLy4 EoS, as shown in Extended Data Figure\,\ref{fig:luminosity}.
In the same figure we also show the upper limits considering the maximum flux across all models (the three atmosphere models and the adopted blackbody model).

\subsection{Magneto-thermal Simulations}
\label{sec: simulations}

Our astrophysical scenario of interest is the long-term evolution of magnetic fields in neutron stars. In the crust of a neutron star, the evolution of the system is governed by two coupled evolution equations: the induction equation and the heat diffusion equation (see the review by \cite{pons2019} for more details), given by:   
\begin{equation}
    \frac{\partial \vec{B}}{\partial t} = -\vec{\nabla}\times \Big[ \eta(T) \vec{\nabla}\times (e^{\nu}\vec{B}) + \frac{c}{4 \pi e n_e}[\vec{\nabla} \times (e^{\nu}\vec{B})] \times \vec{B} \Big]~, 
   \label{eq: induction equation} 
\end{equation}
\begin{eqnarray}
c_{V}(T)  \frac{\partial  \big( T e^{\nu} \big) }{\partial t}  &=&  \vec{\nabla}\cdot (e^{\nu} \hat{\kappa}(T,\vec{B}) \cdot \vec{\nabla}(e^{\nu}T)) + 
 \nonumber\\ 
&&  + e^{2\nu}(Q_{J}(\vec{B},T) - Q_{\nu}(\vec{B},T)) ~,
  \label{eq:thermal_evolution}
\end{eqnarray}
where: $c$ is the speed of light, $e$ the elementary electric charge, $n_e$ the electron number density, $\eta(T)=c^2/(4 \pi \sigma_e(T))$ the magnetic diffusivity (inversely proportional to the electrical conductivity $\sigma_e$), $c_{V}$ the heat capacity per unit volume, $\hat{\kappa}$ the anisotropic thermal conductivity tensor, $Q_J$ and $Q_{\nu}$ the Joule heating rate and neutrino emissivity per unit volume, and $e^\nu$ is the relativistic redshift correction.

The interconnections between the magnetic and thermal evolution equations occur at the microphysical level. On one hand, as the temperature decreases due to neutrino emission, the resistivity of the matter also decreases, leading to increased thermal and electric conductivities. At sufficiently low temperatures, these conductivities become temperature-independent \cite{Aguilera2008}, resulting in a gradual decrease in the Ohmic dissipation rate. On the other hand, as the magnetic field evolves, it affects the thermal conductivity both along and across the magnetic field lines, influencing the local temperature. This, in turn, causes significant variations in the surface temperature distribution $T_s$, which can be observed and constrained through measurements. Simultaneously, the Hall effect drives electric currents towards the crust-core boundary, where the presence of high impurities and pasta phases facilitates a more efficient dissipation of the magnetic field \cite{pons2013}. Consequently, the magnetic energy is converted into Joule heating $Q_J$. While the Hall effect itself does not directly dissipate magnetic energy, it gives rise to small-scale magnetic structures where Ohmic dissipation is enhanced. To a lesser extent, the magnetic field $\boldsymbol{B}$ also influences $c_V$ and $Q_\nu$.

Since the induction equation and the heat diffusion equation are coupled at a microphysical level, they must be supplemented by an EoS, which allows to build the background model of the neutron star solving the Tolman-Oppenheimer-Volkov equations using different models of nuclear EoS at zero temperature (that we take from the online public database CompOSE \url{https://compose.obspm.fr/}). At the temperatures ($T < 10^{10}$\,K) and magnetic field strengths ($B < 10^{16}$\,G) of interest, neutron stars consist of degenerate matter, typically characterized by temperatures lower than the Fermi temperature throughout their entire existence. In this specific temperature range, quantum effects, as dictated by Fermi statistics, overwhelmingly dominate over thermal effects. Therefore, the EoS for neutron stars can be effectively approximated as that of zero temperature, allowing us to largely ignore thermal contributions for most of their lifespan. In particular, here we present results of EoSs at zero temperature, describing both the star crust and the liquid core. This allows us to interpolate the provided tables using different schemes to obtain the relevant quantities, selected by the user. The EoS provides the input to compute at each point of the star the microphysics parameters, e.g., $\eta_b$, $c_V$, $\hat\kappa$, and $Q_\nu$, essential for our simulations. 
In particular, it is important to take into account superfluid and superconductive models for neutrons and protons, respectively, as they significantly impact the cooling timescales through their influence on $c_V$ and $Q_\nu$. 
Additionally, superfluidity and superconductivity activate an additional neutrino emission channel through "Cooper pair breaking and formation" processes. This channel is triggered due to the continuous formation and breaking of Cooper pairs. The effect of the superfluidity and superconductivity, namely the suppression of specific heats, the creation of an additional neutrino channel, and the suppression of neutrino emissivity, collectively influence the cooling process. While the suppression of neutrino emissivity slows down cooling, the other two effects accelerate it. Overall, unless extremely unusual choices are made for the gap models in the neutron star, cooling is typically accelerated. Since the primary objective of this work is not to explore the vast parameter space encompassing superfluidity models, detailed discussions concerning their impact on cooling models and in particular on enhanced cooling are deferred to dedicated studies (e.g., \cite{yakovlev1999, yakovlev2001, page2006,Aguilera2008,page2009,potekhin2015}).
In this study, we utilise the superfluid gap proposed by \cite{ho2015}. 
Various additional microphysical ingredients play a crucial role in the interior of a neutron star. Notably, the EoS and superfluid models have a significant effect on the cooling process as already mentioned, but they play a comparatively smaller role in the evolution of the magnetic field, especially when compared to the initial topology chosen for the system.

For a comprehensive computation of the different microphysics ingredients (e.g., neutrino emissivity, conductivities, magnetic diffusivity, etc) required for the heat diffusion equation and the induction equation in our simulations, we refer to the review by \cite{potekhin2015} and exploring the publicly available codes developed by Alexander Potekhin (\url{http://www.ioffe.ru/astro/conduct/}). These resources offer valuable insights and tools for studying the intricate physics governing neutron stars.

To solve the induction equation (eq.\,\ref{eq: induction equation}), we consider the magnetic field confined in the crust of a neutron star assuming perfect conductor boundary condition at the crust-core interface and potential magnetic boundary conditions (current-free magnetosphere) at the outer numerical boundary, placed at density $\rho=10^{10}$ g~cm$^{-3}$. Moreover, in this study, we focus solely on the axisymmetric evolution and do not consider the generalisation to 3D, as performed in the recently developed \textit{MATINS} code for coupled magneto-thermal evolution in isolated neutron stars \cite{dehman2022,dehman2023c}.

We limit our analysis to solving the heat diffusion equation (assuming an initial temperature value of $10^{10}$\,K) solely in the crust of the neutron star. That is due to the fact that the core of a neutron star becomes isothermal a few decades after formation, due to its high thermal conductivity. Meanwhile, the low-density region (envelope and atmosphere) quickly reaches radiative equilibrium, establishing a significant difference in thermal relaxation timescales compared to the interior (crust and core). This discrepancy makes it computationally expensive to perform cooling simulations on a numerical grid encompassing all layers up to the star's surface. Instead, for the outer layer, we rely on a separate calculation of stationary envelope models to derive a functional relationship between the surface temperature $T_s$ (determining the radiation flux) and the temperature $T_b$ at the crust/envelope boundary. This $T_s$-$T_b$ relation serves as the outer boundary condition for the heat transfer equation.

To account for surface radiation, we adopt a blackbody emission model. Given our focus on explaining the luminosity of faint and young objects, we specifically adopt the most updated iron envelope \cite{potekhin2015}. Light element envelope models, in contrast, result in approximately one order of magnitude brighter luminosity. For a more comprehensive examination of how envelope models with different compositions influence the cooling process, we direct the reader to \cite{dehman2023b}.

In this study, we employ the latest version of the 2D magneto-thermal code \cite{vigano2021} to model nuclear matter. Additionally, to model hyperon matter in the core of the star, we use the code in \cite{anzuini2021,anzuini2022}, which is also based on \cite{vigano2021}, but has been appropriately modified to calculate the magneto-thermal evolution in stars containing both nuclear and hyperon matter. This code smoothly matches the GM1A and GM1B EoSs \cite{gusakov2014} in the core, with the SLy4 EoS \cite{douchin2001} in the crust while also considering the influence of hyperon matter on the star's microphysics.
One crucial modification made in \cite{anzuini2021, anzuini2022}, in comparison to \cite{vigano2021}, is the superfluid gap model, which allows accurate descriptions of stars containing hyperons. Furthermore, the most significant effect of this study arises from the inclusion of the hyperon Direct Urca channel and the Cooper pair of hyperons. These two neutrino cooling channels, especially the hyperon Direct Urca one, lead to enhanced cooling in neutron stars at a young age. 

Our simulations encompass various neutron star background models, considering different masses $M=1.4, 1.6,$ and $1.8$ M$_\odot$, along with diverse nuclear EoSs. We selected three EoSs for our investigation: BSK24 \cite{pearson2018}, SLy4 EoS \cite{douchin2001}, and GM1A EoS \cite{gusakov2014} matched with the SLy4 EoS in the crust \cite{anzuini2021,anzuini2022}.  Through magneto-thermal simulations, we observed distinctive cooling behaviours. Specifically, the SLy4 EoS \cite{douchin2001} demonstrated no activation of enhanced cooling. Conversely, the BSK24 EoS \cite{pearson2018} exhibited enhanced cooling, triggered at a Direct Urca threshold density of $8.25 \times 10^{14}$\,g\,cm$^{-3}$. It is noteworthy that the latter EoS attains a maximum mass of $2.279$\,M$_\odot$ and a corresponding central density of $2.26 \times 10^{15}$\,g\,cm$^{-3}$. In the case of the GM1A EoS \cite{gusakov2014}, our analysis revealed consistent activation of enhanced cooling across all examined masses, starting from a mass of $1.4$\,M$_\odot$ and a density of $5.949 \times 10^{14}$\,g\,cm$^{-3}$. This particular EoS reaches a maximum mass of $1.994$\,M$_\odot$ and a corresponding central density of $2\times 10^{15}$\,g\,cm$^{-3}$.

Furthermore, we explore a wide range of initial magnetic field values for the surface dipolar field, ranging from $1\times10^{12}$\,G to $7\times10^{13}$\,G at the pole. Our simulations are restricted to this range in accordance with our objective to analyze pulsars with magnetic fields ranging from $10^{12}$\,G to a few $10^{13}$\,G, as confirmed by observational measurements. To maintain focus on the cold neutron star scenario, we intentionally excluded the inclusion of a toroidal field in our simulations. This decision was made to prevent Joule heating and to avoid scenarios where the dissipation of the magnetic field in the highly resistive crust might obscure the effects of enhanced cooling mechanisms in stars with magnetic fields exceeding $10^{14}$\,G \cite{Aguilera2008}. In Supplementary Data Figure 2 we show the same simulations as for Figure\ref{fig:cooling} but showing the period and period derivative evolution.



\subsection{On the effect of choosing different envelopes} \label{sec:envelopes}

In this section we explain the reasoning behind some of the assumptions and/or simplifications we make in this study, in particular concerning the neutron star envelope, and the ages and distances we use for the three cold neutron stars we present.

Neutron star envelopes might be composed by a variety of elements, from the lightest ones as Hydrogen or Helium, to heavy envelopes as Iron for example. The envelope composition might differ from source to source depending on its exact evolutionary history, and they are known to show different cooling evolution \cite{dehman2023b}. Cooling simulations by many authors show how heavy envelopes systematically result in cooler temperatures at younger ages. We have chosen to use here only the simulations using the heavy envelope since we aimed at the modelling of very faint sources, hence we assumed the most extreme case. If these objects would possess a light envelope, the need of enhanced cooling would be even more pronounced. 

On the other hand, while initially considering also the pulsar PSR\,B0656$+$14 among the extremely cold sources, while performing simulations using also light envelopes \cite{dehman2023b}, we saw that at such older ages ($>10^4$\,yr) the cooling curves might behave differently (see Extended Data Figure\,\ref{fig:envelope-comp}). In particular, using light envelopes, even EoSs not activating enhanced cooling processes might explain this object.

In Extended Data Figure\,\ref{fig:envelope-comp} we also show simulations for a 2 M$_\odot$ source, showing how even for this extreme mass the SLy4 does not activate any enhanced cooling mechanism.

\subsection{Statistical analysis}\label{sec:statistical}

To constrain the nature of each source, we start comparing their observed parameters, $L_{\rm th}$, $P$ and $\dot{P}$, against the tracks of the 81 simulations in the same 3D parameter space (see Extended Data Figure\,\ref{fig:3dplots}), each containing 128 points corresponding to various times from the formation of the NS up to ${\sim}100\rm\,kyr$ age ($0, 1.3, 1.5, 1.8, \ldots, 81200, 97400\,{\rm yr}$). Ideally, the most probable parameters (EoS, mass, and $B_p$) of a given source are those of the simulated neutron star sharing the same observed features. However, due to the finiteness of the simulations, and the uncertainty on the luminosity (we ignore the uncertainties on $P$ and $\dot{P}$ since the relative uncertainty on $L_{\rm th}$ is significantly higher), there is no simulation passing exactly through the positions of the sources in the feature space. Additionally, multiple simulations may be found in the vicinity of the sources (see Extended Data Figure\,\ref{fig:3dplots}). Consequently, we opt for a machine learning model that given the observational and simulation data, can identify the most probable simulations, and as a consequence, the posterior distribution of the parameters, namely, the EoS, mass and $B_p$. We tried with two approaches, deep-learning and classification, explained in detail in the next paragraphs.

\subsection{The deep learning approach}
\label{txt:deeplearning}

The simulation parameters are either categorical (EoS) or continuous (mass and $B_p$). If the categorical and continuous parameters could be constrained by mutually exclusive sets of features (e.g., if the EoS could be constrained only by $L_{\rm th}$, while mass and $B_p$ only by $P$ and $\dot{P}$), then their estimation could rely on independent classification and regression models on the independent sets of inputs and outputs. However, this is not the case in the simulations: we need to employ a machine learning approach that performs both classification and regression simultaneously.
We used a neural network that learns the parameters (EoS, mass and $B_p$) given a specific point in the feature space ($L_{\rm X}$, $P$ and $\dot{P}$), trained on the simulation data.
Specifically, using \textsc{TensorFlow} \cite{tensorflow} we constructed a multilayer perceptron (MLP) neural network that predicts the parameters, with a loss function being the sum of the loss for the EoS, and the loss for the values of mass and $B_p$. The architecture is summarised as an input layer of size 3 (for the three features), fully-connected hidden layers (with rectified linear unit activation functions), and two output layers connected to the last hidden layer: (i) the classification output layer of size 3 and softmax activation function (for the classification probabilities for BSK24, SLy4, SLy4+GM1A), and the (ii) regression output layer of size 2 and linear activation function (for the mass and $B_p$). We tried different numbers and sizes of hidden layers, loss functions (e.g., mean squared error of the regression output, and cross entropy for the classification output). We found that the models were converging slowly (up to 10000 epochs) with poor results: the best accuracy in predicting the EoS class was ${\sim}65\%$ which is not significant with respect to a \emph{dummy} classifier (33\% accuracy if randomly assigning one of the three EoSs) and the classification approach below (90\% accuracy). Increasing the number of simulations would aid the neural network in learning the feature space, however owning to their computational complexity, we employed the classification approach.


\subsection{The classification approach}
\label{txt:machine}

Interestingly, classification can be seen as \lq{}discretised\rq{} regression: instead of estimating the mass of a NS, we can classify it into distinct mass classes, 1.4, 1.6 and $1.8\,\rm M_\odot$. However, training three classifiers for EoS, mass and $B_p$ is not optimal since this ignores the interplay between the parameters in the evolution of neutron stars imprinted in the feature space. In our case, each simulation corresponds to a combination of EoS, mass and $B_p$ classes, and therefore there is only one class: the \emph{simulation class} which can be modelled as the ID of the simulation (i.e., $k$ corresponding to the $k$-th simulation). This is a simple classification task which can be easily carried out with standard, well-understood machine learning classifiers that are easily trained on small data sets. Using a probabilistic classification algorithm to predict the simulation class, we can predict the classification probability of each simulation for each observed source. Then, the posterior of each parameter to have a specific value is simply the sum of all the classification probabilities of the simulations sharing the same value. For example, the posterior of the EoS of a source being BSK24 is the sum of the classification probabilities of all the simulations (classes) for which the EoS is BSK24, predicted on the features of the source:
\begin{equation}
P(\text{BSK24}) = \sum_{k=1}^{n} P(\text{BSK24}=\text{EoS}_k) \pi_k
\end{equation}
where $P(\text{BSK24}=EoS_k)$ is either 1 or 0, denoting whether the BSK24 EoS was used in the $k$-th simulation, $\pi_k$ is the prior on the class which in our case is the classification probability of the $k$-th simulation when predicting the class of the observation, and $n$ is the total number of simulations. The same approach is used for the continuous parameters as well (mass and $B_p$), with the output being still the marginal probabilities at the distinct values used in the simulations.

We stress that one should be careful in the interpretation of the various classification metrics (e.g., accuracy). The number of simulation classes depends on the choices for the range and resolution in the initial conditions, which are generally restricted due to technical and modelling difficulties (e.g., computational cost, available EoS models, etc.) Ideally, a large number of simulations could be run, leading to a paradox: due to the continuous nature of the mass and $B_p$, the classification probability would approach 0 even if the \lq{}correct\rq{} model is present in the training. Consequently, the absolute scale of the accuracy of the trained classifier is not an estimate of the performance of the methodology. Conversely, the relative accuracy between different algorithms (or hyperparameters) measures their relative ability to learn the feature space given the simulation choices and the observational uncertainties.

\subsection{Selection of classifier and hyperparameters}

To select the classification algorithm, we design a cross-validation test bed. We consider 8 different classifiers offered by the \textsc{scikit-learn} package \cite{sklearn}: k-nearest neighbour, random forest, decision tree, logistic, support vector (SVC), nu-SVC, MLP, and Gaussian process classifier, and multiple hyperparameter choices for each (3-10 different values for a key hyperparameter such as $k$ for the $k$-nearest neighbour, or the kernel for the Gaussian process classifier).

We set aside $^1/_6$ of the data as a test data set that will be used to estimate the accuracy of the classifier with the best hyperparameters. The remaining data set is separated in 5 folds of equal fractions ($^1/_6$ of the original data). A 5-fold cross-validation approach is adopted to measure the accuracy of the different classifiers and hyperparameter choices. However, the test samples fall very close to the training samples since they follow distinct curves in the feature space. Consequently, most classifiers will exhibit very high performance which does not reflect the accuracy when applied in real data which are subject to measurement uncertainties. For this reason, we \lq{}disturb\rq{} the cross-validation samples by adding Gaussian noise in the decimal logarithm of the luminosity of the simulations (we ignore the uncertainties on $P$ and $\dot{P}$ since they are negligible), to simulate the presence of uncertainty. We train, cross-validate, and test the classifiers at six different scales for the uncertainty: 0 (no disturbance), 0.05, 0.10, 0.15, 0.2, 0.25, and 0.30 dex, a range that includes the uncertainty on $L_{\rm th}$ in our sources (close to $0.2\rm\,dex$). 

For each classifier and $L_{\rm th}$ uncertainty level, we use the cross-validation technique to optimise for the hyperparameters. Then, using the test data set we measure the accuracy score, i.e., the fraction of test data points that the algorithm was able to match to their original track. 
Trying all classification algorithms initially, we found that the $k$-nearest neighbours and random forest classifiers presented the highest accuracy scores. Additionally, they are computationally efficient during both training and prediction. For this reason, we focus on these two classification methods from now on.
In the top left panel of Supplementary Data Figure 3, with solid lines, we show the accuracy of the $k$-nearest neighbour and the random forest classifier, as a function of the $L_{\rm th}$ uncertainty level. Both algorithms perform equally well. For the prediction of the properties of the observed source, we select the random forest classifier because it performs better at high uncertainties (${\gtrsim}0.2\,\rm dex$). Additionally, the random forest classifier has two attractive properties: it's not sensitive to the scale of the features, and it is intrinsically a probabilistic algorithm.

Finally, since we are interested in the ability to predict the physical properties of the pulsars, we measure the marginal accuracy of the classifiers, i.e., the ability to predict them independently. For example, if a test data point corresponds to a model with $M{=}1.4\rm\,M_\odot$, does the predicted model have the same mass (no matter what the EoS or $B_p$)? In the top right, bottom left and bottom right panels of Supplementary Data Figure 3, we show the marginalised accuracy for the EoS, mass and $B_p$, respectively (with solid lines). We find that the magnetic field is easier to be learned (${>}95\%$ accuracy), while the EoS and mass are sometimes mismatched ($43-95\%$ accuracy), especially at high uncertainty levels. The fact that the accuracy score is not 100\% even with no added noise in the luminosity, reflects the degeneracy between the models: as it can be seen in Extended Data Figure\,\ref{fig:3dplots}, the tracks of the different simulations often occupy the same regions of the feature space. Here, we remind that the accuracy score should be used only for comparisons of algorithms, and not as a measure of performance of the methodology.

\subsection{Predictions accounting for the uncertainty on the luminosity}
\label{txt:asymmetric}

We use the random forest classifier that has been optimised for the level of $L_{\rm th}$ uncertainty in our observations, retrained using all the available simulation data (without separation to training, validation, and test data sets; \cite{Tsamardinos22}). However, the uncertainty of the luminosity is not used directly during prediction. To take into account the observational uncertainty of a given source's luminosity, we use a Monte Carlo approach: we sample the error distribution of the luminosity 100,000 times, predict the properties of the sources, and sum the results. To sample the error distribution of the luminosity, we use Monte Carlo uncertainty propagation: we model the error distributions of the flux and the distance, draw samples from them, and calculate the luminosity samples. 
This is to avoid standard uncertainty propagation for three reasons: (i) we have high relative uncertainties in the quantities involved; (ii) the flux and distance confidence intervals are not symmetric (especially in the case of the distance of \psrb); (iii) the classifier operates in log-space where even symmetric error bars are transformed into asymmetric ones. We note that applying the standard uncertainty propagation formula by averaging the low and high error bars resulted to ${\sim}20\%$ differences in the resulting classification probabilities with respect to the Monte Carlo uncertainty propagation using asymmetric error distributions as outlined below.

First, we represent the flux and distance uncertainties using the binormal distribution which is flexible enough to represent asymmetric distributions. A binormal distribution's probability density function (PDF) is the result of stitching together the opposite halves of two distinct normal distributions with the same mean value (which acts as the mode of the new distribution) but different standard deviations \cite{Wallis14}. Consequently, we represent each flux and distance measurement with a binormal distribution by adopting the measured value as the mode, and fitting for the two standard deviations such that the reported confidence intervals are matching those of the binormal distribution. In Supplementary Data Figure 4 we show the PDFs of the constructed binormal distributions (in the form of histograms of their samples) for the flux and distance (left and middle panel, respectively) of the two sources for which we apply the classification, as well as the derived luminosity error distribution.

We predict the probability of the 81 simulations, 100,000 times (for each sample from the $L_{\rm th}$ distribution) for each source. By summing up the 100,000 results, we find the classification probability for each model. In the top panel of Supplementary Data Table~2 we show the models with the highest classification probabilities, while in Figure~\ref{fig:statistical} we visualise all the results. In addition, in the table, we report the marginalised probability for the EoS, mass and $B_p$ of each source, which are also shown in Supplementary Data Figure 5.

\subsection{Accounting for the age information}

Since the simulations track the evolution of the properties of the pulsars in time, if the real age of the source is known, it provides an additional constraint. We add the time as another input variable, making the feature space four-dimensional, and then repeat the above analysis: (i) optimise the two classification algorithms for different luminosity uncertainty scales (see dotted lines in Supplementary Data Figure 3); (ii) select the random forest classifier optimised for the $0.2\rm\,dex$ uncertainty scale (it outperforms the $k$-nearest neighbour classifier at uncertainty scales ${>}0.1\rm\,dex$); (iii) predict the parameters of the two sources using the age estimates in Extended Data Table~\ref{tab:par}.

Having a 4D feature space, it is impossible to visually inspect its coverage by the simulation tracks. Instead, we ensured that the observed sources fall within the range of the simulation evolutionary tracks by confirming that their positions are inside a simplex of the Delaunay hypertetrahedralisation of the simulation data.

The predictions of the 4D classifier are shown in Extended Data Table~\ref{tab:classprob4D} and Supplementary Data Figure 5 (with red rectangles).
These results are for the most part consistent with the 3D classifier. For both sources, the 3D and 4D classifiers indicate the same most probable value for the $B_{p}$. For \psra, the most probable values for the EoS and mass are the same, while for \psra they differ, but not significantly (BSK24 and GM1A EoS, and 1.6 and 1.8 $\rm M_\odot$ masses have high marginal posteriors ${>}25\%$).

We notice that the posteriors in the 4D case are less ``peaky''.
This is in contrast to our expectation that adding another feature (age) would create larger contrast between the marginal probabilities (effectively lowering the entropy of information). 
Given the small number of EoSs in our simulations, at this stage, we do not consider our method well suited to constrain the EoS itself. Nevertheless, for both sources, our statistical analysis do shows that the SLy4 EoS and the $1.4\,\rm M_\odot$ mass scenario are found to be highly improbable with or without considering the age information.

\end{methods}


\makeatletter
\apptocmd{\thebibliography}{\global\c@NAT@ctr 43\relax}{}{}
\makeatother

\renewcommand\thefigure{\arabic{figure}} 
\setcounter{figure}{0}
\renewcommand{\figurename}{Extended Data Figure}
\renewcommand{\tablename}{Extended Data Table}

\begin{figure}[htp!]
	\includegraphics[width=1.0\columnwidth]{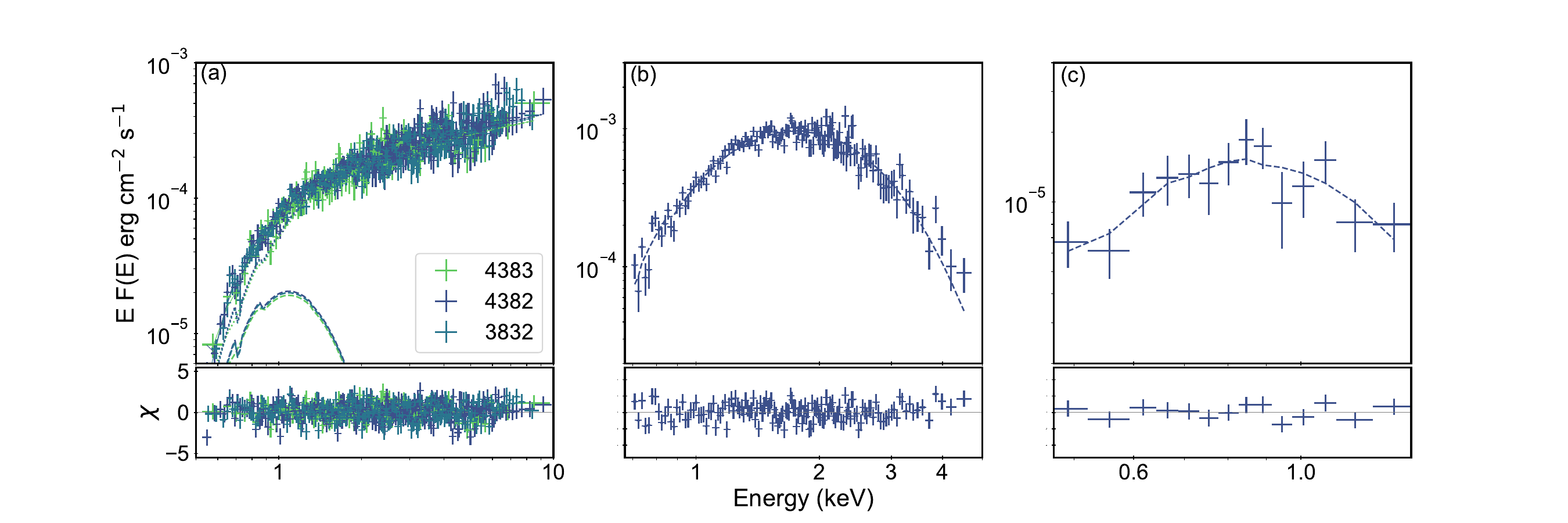}
    \caption{Energy spectra, best-fit model and residuals for the three sources used in this manuscript, i.e., (a): \psra 
 (using three different observations labelled with the relative ID; see Extended Data Table \ref{tab:log}), (b): \psrb , (c): \velajr . Different line styles were adopted to distinguish between the different components, i.e. dotted for \textsc{powerlaw} and dashed for \textsc{bbodyrad}.}
    \label{fig:spec}
\end{figure}

\newpage

\begin{figure}[htp!]
    \includegraphics[width=0.33\columnwidth]{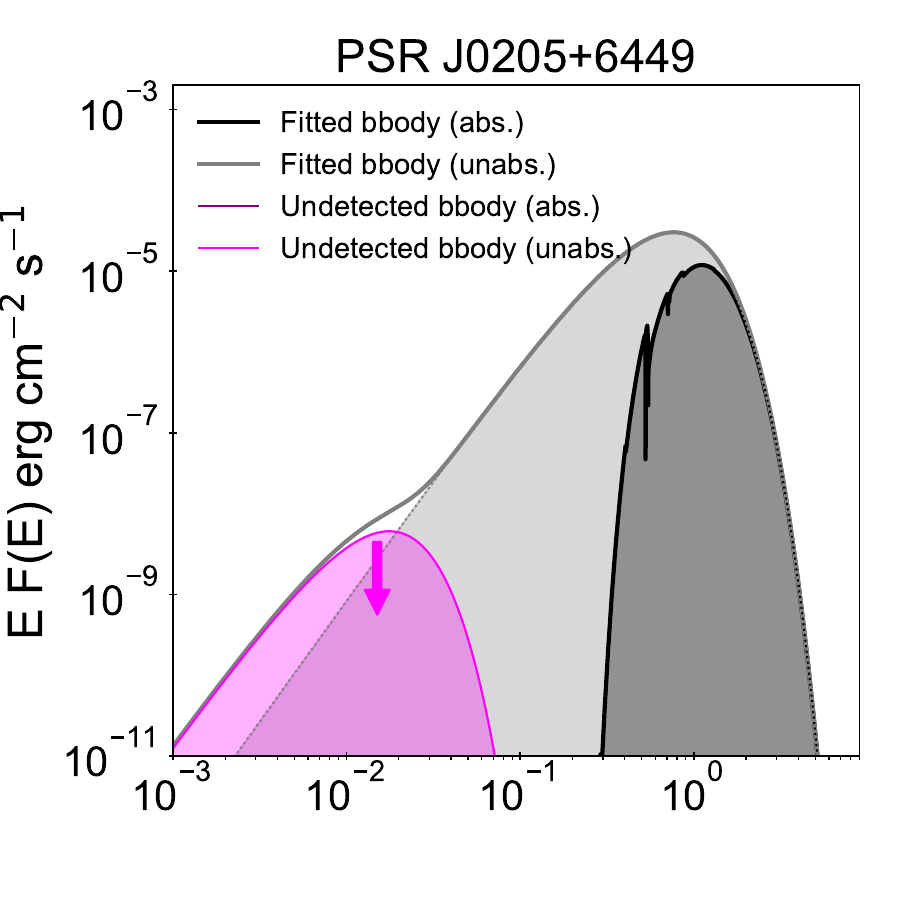}
    \includegraphics[width=0.33\columnwidth]{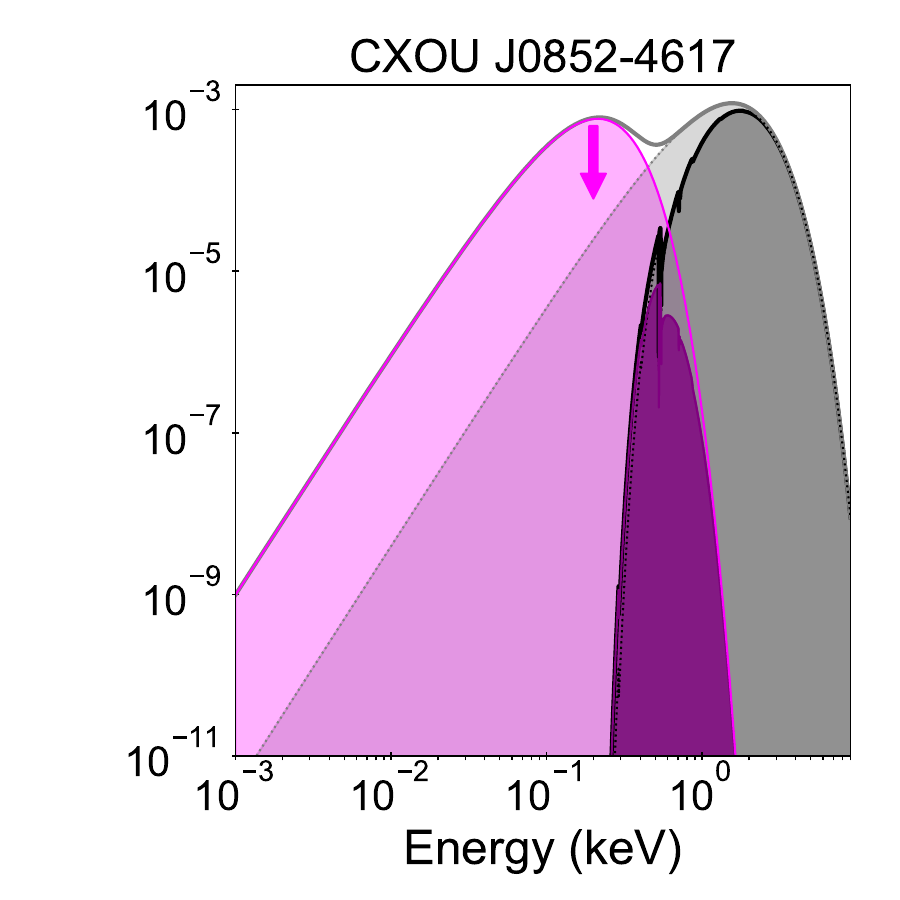}
	\includegraphics[width=0.33\columnwidth]{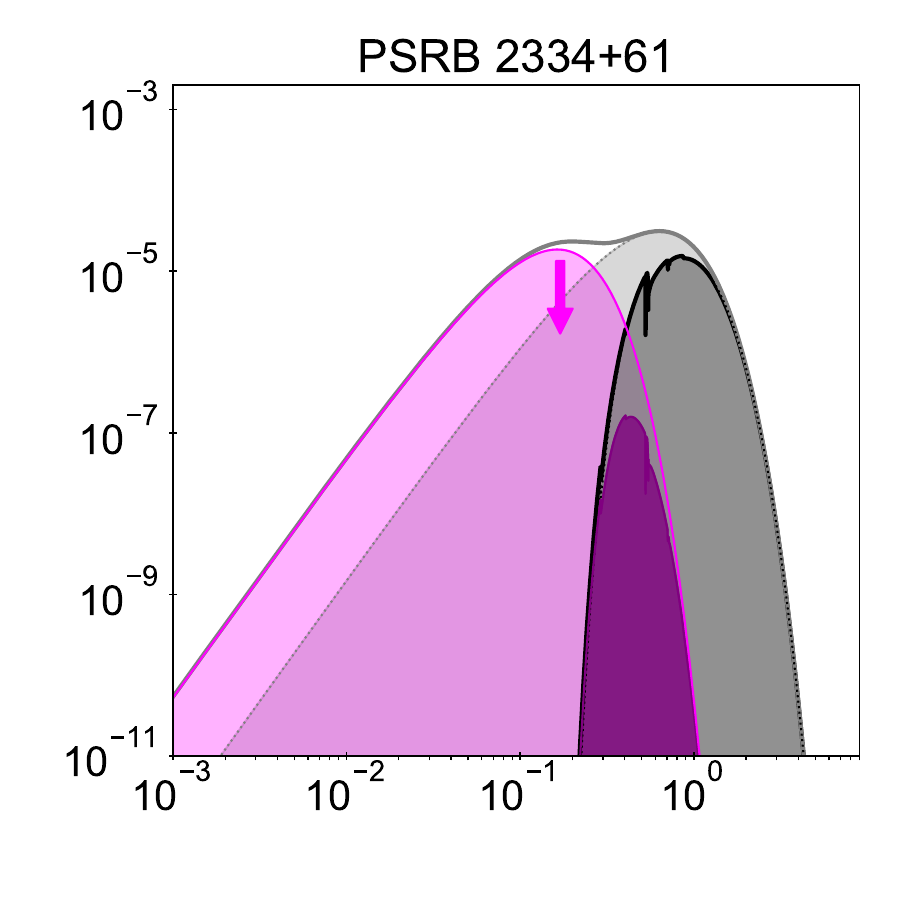}

    \caption{Comparison between best-fit models of the thermal emission in the three sources used in this work and the cold, undetected blackbody components which may possibly be emitted by the whole NS surface but hidden out by interstellar absorption (see Extended Data Table \ref{tab:fit_results}). The absorbed components are displayed with bolder lines and colors to distinguish them from the unabsorbed components.}
    \label{fig:coldbb}
\end{figure}

\newpage

\begin{table*}
\caption{Timing properties and age estimates}
\begin{center}
\begin{tabular}{l c c c }
\hline
\hline
& \psra & \psrb & \velajr \\
\hline
Class & RPP & RPP & CCO \\
assoc./nick & 3C 58 & SNR G114.3+0.3 & SNR Vela Jr \\
$d$ [kpc] & $2.0\pm0.3$\textsuperscript{a} & $3.1^{+0.2}_{-1.0}$\textsuperscript{b} & $0.5-1$\textsuperscript{c} \\
$P$ [s]& 0.06565923(2)$\textsuperscript{d}$ & 0.495228(3)$\textsuperscript{e}$ & - \\
$\dot{P}$ [$\times$10$^{-13}$ s/s] & 1.935$\pm$0.003$\textsuperscript{d}$ & 1.9098$\pm$0.0003$\textsuperscript{f}$ & - \\
$\dot{E}_{rot}^\dagger$ [$\times$10$^{34}$ erg/s] & 2666 & 6.1 &- \\ 
$B_p^\ddagger$ [$\times$10$^{12}$ G] & 7.2 & 19.7 & - \\
$\tau_c$ [kyr] & 58.8 & 41.1 & - \\
$\tau_r$ [kyr] & 0.80$^{g,*}$ & 7.7$\pm$0.8$^{h,**}$ & 2.5-5.0$^c$ \\
\hline
\hline
\end{tabular}
\label{tab:par}
\end{center}
$\dagger$: Rotational energy loss, $\dot{E}_{rot}=3.9\times 10^{46}\dot{P}/P^3$ erg/s;\\ $\ddagger$: Magnetic field strength at the pole, assuming that rotational energy losses are dominated by dipolar magnetic torques, $B_p=6.4\times 10^{19}(P\dot{P})^{1/2}$~G. \\
*: A conservative 10\% error was adopted only in the analysis and Figure\,\ref{fig:cooling}, despite the date was inferred for the actual detection of the supernova.\\
**: A conservative 10\% error was adopted in this case to account for any systematics in the SNR age characterization reported in \cite{YarUyaniker2004}.\\
Distance and age references: 
a:\cite{Ranasinghe2022}, b:\cite{McGowan2006}, 
 c: \cite{Allen2015}, d:\cite{Murray2002}, e:\cite{Dewey1985}; f:\cite{Dewey1988}; g:\cite{Kothes2013}, 
h: \cite{YarUyaniker2004}. 
\end{table*}

\newpage

\begin{table*}
\caption{Spectral analysis results \label{tab:fit_results}}
\begin{center}
\begin{tabular}{ l c c c }
\hline 
\hline
\bf{Parameters} &  \psra & \psrb & CXOU J0852 \\
\hline
Model & BB+PL & BB & BB \\ 
$N_{\rm H}$ {($\times$10$^{22}$ cm$^{-2}$)} & 0.49$^{+0.08}_{-0.05}$ & 0.20$^\phi$ & 0.47$\pm$0.04  \\
$\Gamma$ & $1.51^{+0.07}_{-0.06}$ & - & - \\
$kT_{\rm bb}$ (keV) & $0.20^{+0.08}_{-0.04}$ & $0.16\pm$0.01 & 0.397$\pm0.009$ \\
$R_{\rm bb}$ (km) & 0.6$^{+1.5}_{-0.3}$ & 0.9$^{+0.7}_{-0.6}$ & 0.24$\pm$0.08 \\
$\chi^2_\nu$ (d.o.f.) & 0.96 (598) & - & 1.13 (143) \\
C-stat (d.o.f.) & - & 6.46 (12) & -  \\
\hline
$F_{\rm th}$ ($10^{-13}\rm\,erg\,cm^{-2}\,s^{-1}$) 
    & $3.02^{+0.79}_{-0.61}$ 
    & $0.68 \pm 0.10$ 
    & 25.9$^{+1.1}_{-1.0}$ \\
$kT_{\rm cool}$ (keV) & $<4\times10^{-3}$ & $<4\times10^{-2}$ & $<5\times10^{-2}$ \\

\hline
\hline
\end{tabular}
\end{center}
$^\phi$: Kept frozen during the fits.
Quoted errors reflect 90\% confidence levels. See uncertainty propagation analysis in \S\ref{txt:asymmetric} for the $L_{\rm th}$ of two of the sources (\textsuperscript{\dag}). The reported best-fit parameters have been obtained assuming absorption tables by \cite{Wilms2000}.
\end{table*}

\newpage
\begin{table*}
\caption{Log of the \xmm and \chandra observations used in this work.}
\centering
\begin{tabular}{ccccccc}
\hline
\hline
Source & Date Obs.     & Obs.ID      &  Satellite         & Exp. [ks] & Cts. [$10^3$] & Refs. \\
\hline      
    \hline
\multirow{3}{*}{\psra} &    2003-04-22 &    4383    &   \multirow{3}{*}{\chandra} &  38.7   & 4.82$\pm$0.07    &  \cite{Slane2004}    \\
&   2003-04-23 &    4382    &                             &  167.4  & 22.05$\pm$0.15    &  \cite{Slane2004}  \\                   
  &  2003-04-26 &    3832    &                             &  135.8  & 17.72$\pm$0.13    & \cite{Slane2004}    \\
    \hline
 \psrb &   2004-02-12 &    0204070201  & XMM &  26.8 & 0.32$\pm$0.02  & \cite{McGowan2006} \\
    \hline
 \velajr &   2001-09-17 & 1034 & \chandra & 31.4 & 9.5$\pm$0.1 & \cite{Kargaltsev2002} \\
\hline
\hline
\end{tabular}
\label{tab:log}
\end{table*}

\newpage

\begin{table*}
\caption{Spectral analysis for the atmosphere model \textsc{nsmaxg}. \label{tab:fit_atmo_results}}
\begin{center}
\begin{tabular}{ l c c c }
\hline 
\hline
Parameters &  \psra & \psrb & CXOU J0852 \\
\hline
 \multicolumn{4}{c}{Chemical composition: H} \\ 
\hline
$kT_{\rm eff}$ (keV) & 0.19$\pm$0.04 & 0.12$\pm$0.03 & 0.31$\pm$0.01\\
$N_{\rm nsmaxg}^{\psi}$ ($10^{-3}$) & 3.0$^{+3.0}_{-1.0}$ & 70$^{+70}_{-30}$ & 2.1$^{+0.6}_{-0.4}$\\  
$F_{\rm th}$ ($10^{-13}\rm\,erg\,cm^{-2}\,s^{-1}$) 
    & $0.70\pm0.10$ 
    & $0.80\pm0.10$ 
    & $27.0\pm0.5$ \\
    \hline
 \multicolumn{4}{c}{Chemical composition: C} \\ 
 \hline
$kT_{\rm eff}$ (keV) & 0.20$\pm$0.03 & 0.10$\pm$0.01 \\
$N_{\rm nsmaxg}^{\psi}$ ($10^{-3}$) & 2.0$^{+3.0}_{-1.0}$ & 20$^{+4}_{-1}$\\  
$F_{\rm th}$ ($10^{-13}\rm\,erg\,cm^{-2}\,s^{-1}$) 
    & $0.80\pm0.20$ 
    & 0.56$\pm$0.07
    & - \\
    \hline
 \multicolumn{4}{c}{Chemical composition: O} \\  
 \hline
$kT_{\rm eff}$ (keV) & 0.36$\pm$0.13 & 0.25$\pm$0.02 \\
$N_{\rm nsmaxg}^{\psi}$ ($10^{-3}$) & 0.14$^{+0.12}_{-0.06}$ & 31$^{+14}_{-10}$ \\  
$F_{\rm th}$ ($10^{-13}\rm\,erg\,cm^{-2}\,s^{-1}$) 
    & $0.70^{+0.12}_{-0.05}$
    & $0.68\pm0.10$
    & - \\
\hline
\hline
\end{tabular}
\end{center}
All fits were performed assuming $B=10^{12}$ G and fixing the distance to the values reported in Extended Data Table \ref{tab:par}; $^\phi$: Kept frozen during the fits. $^{\psi}$: $N_{\rm nsmaxg}=(R_{\rm em}/R_{\rm NS})^2$. 
Quoted errors reflect 90\% confidence levels.
\end{table*}

\newpage

\begin{figure}[htp!]
	\includegraphics[width=\columnwidth]{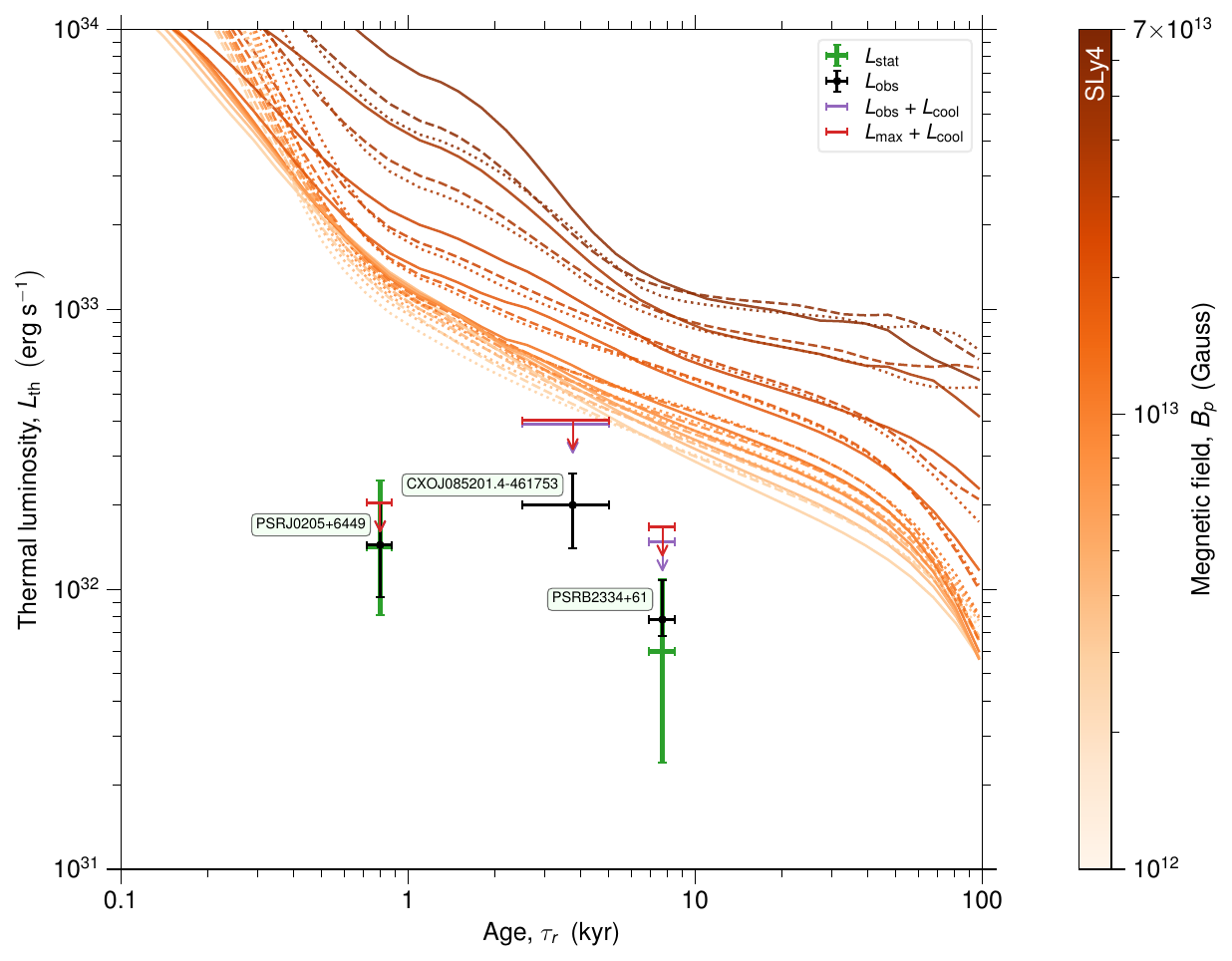}
    \caption{Comparison between the luminosities obtained in this work (see Supplementary Data Table 1) and theoretical cooling curves simulated adopting SLy4 as EoS. The curves are stylised following the same prescriptions as in Fig. 1. For each of the three sources we show $L_{\rm obs}$ ($L_{\rm stat}$) with black (green) error bars. Upper limits on the luminosity of each sources, calculated as $L_{\rm obs}+\Delta L_{\rm obs}+L_{\rm cool}$ (with $\Delta L_{\rm obs}$ the 1-$\sigma$ error on $L_{\rm obs}$), are displayed with purple vertical arrows. An alternative version of the upper limits, using the highest flux from all models (three atmosphere models and the adopted blackbody model, see Extended Data Tables 3-4, are shown with red vertical arrows}.
    \label{fig:luminosity}
\end{figure}


\newpage

\begin{figure}[htp!]
	\includegraphics[width=1.0\columnwidth]{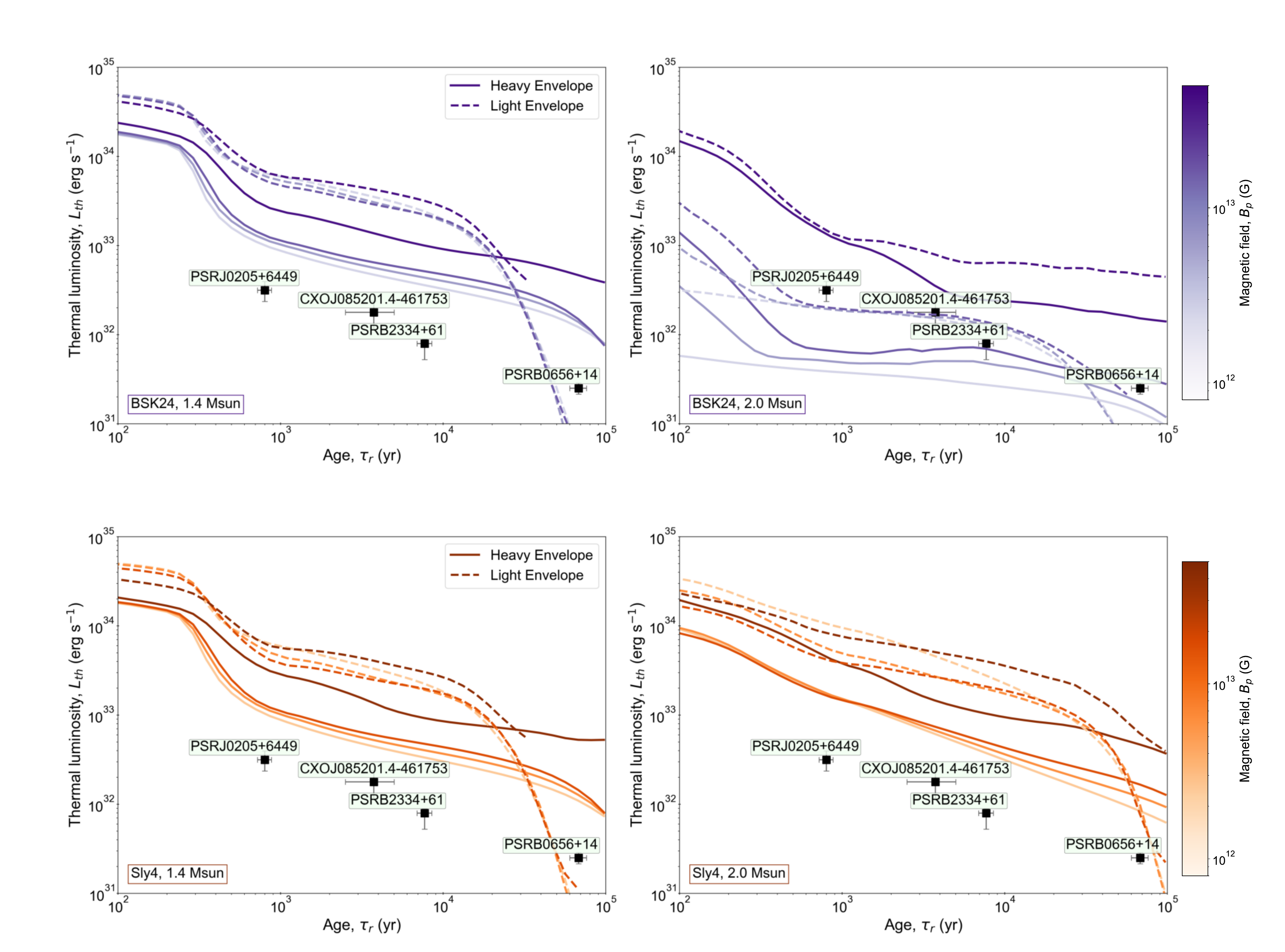}
    \caption{Comparison between cooling tracks produced using heavy (solid lines) and light (dashed lines) envelopes. The curves in the top (bottom) panels were produced using BSK24 (SLy4) and are colored with purple (orange). Different shades of the same colours are used to identify different values of the magnetic fields, as shown in the colour bars. The observed values of \psra , \psrb , \velajr and PSR B0656+14 are reported in each panel.}
    \label{fig:envelope-comp}
\end{figure}

\newpage

\begin{figure}[htp!]
    \includegraphics[width=\columnwidth]{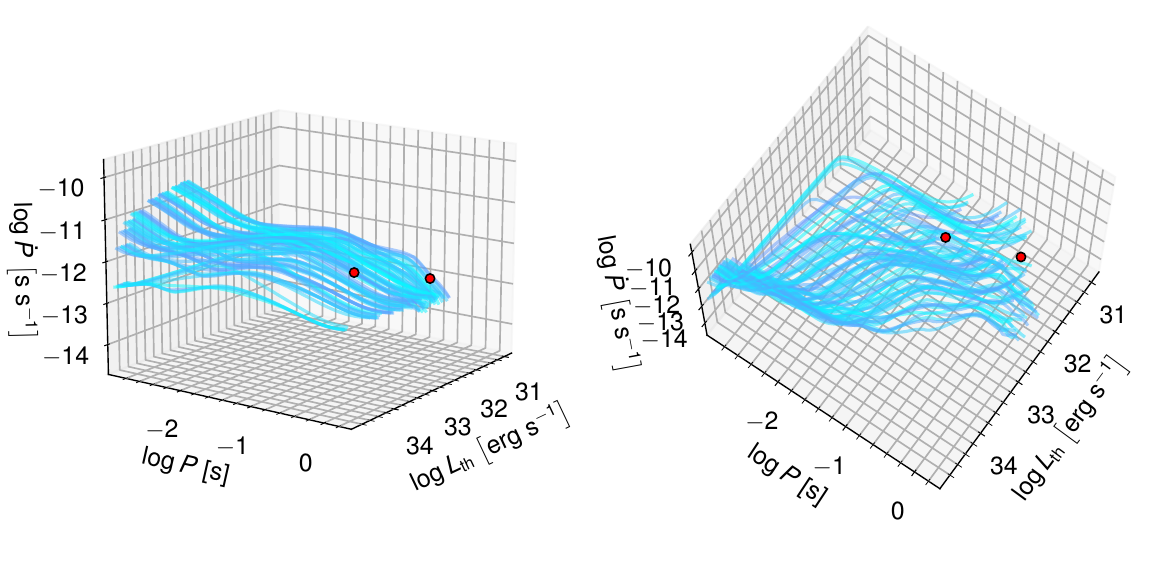}
    \caption{The tracks of our 3D simulations in the $P{-}\dot{P}{-}L_{\rm th}$ space in a time span of $100\rm\,kyr$ (cyan lines) from two different view angles (left and right panels). The positions of \psra and \psrb are denoted with red discs.}
    \label{fig:3dplots}
\end{figure}


\newpage

\begin{table*}
    \centering
    \begingroup
    \footnotesize
    \setlength{\tabcolsep}{8pt} 
    \renewcommand{\arraystretch}{0.7} 
    \begin{tabular}{|l|c||l|c|}
    \hline

\multicolumn{2}{|c||}{PSRB2334+61}  &  \multicolumn{2}{c|}{PSRJ0205+6449} \\
 \hline \multicolumn{4}{|c|}{\textsc{I. Five best models and their classification probabilities}} \\ \hline 
GM1A, $1.6\,M_\odot$, $2{\times}10^{13}$G  &  0.1980  &  BSK24, $1.6\,M_\odot$, $7{\times}10^{12}$G  &  0.1203 \\
BSK24, $1.8\,M_\odot$, $1{\times}10^{13}$G  &  0.1641  &  BSK24, $1.8\,M_\odot$, $2{\times}10^{13}$G  &  0.0857 \\
BSK24, $1.8\,M_\odot$, $2{\times}10^{13}$G  &  0.1087  &  BSK24, $1.6\,M_\odot$, $5{\times}10^{12}$G  &  0.0755 \\
BSK24, $1.8\,M_\odot$, $3{\times}10^{13}$G  &  0.0651  &  BSK24, $1.6\,M_\odot$, $1{\times}10^{13}$G  &  0.0723 \\
BSK24, $1.8\,M_\odot$, $7{\times}10^{12}$G  &  0.0629  &  BSK24, $1.6\,M_\odot$, $3{\times}10^{12}$G  &  0.0696 \\
 \hline \multicolumn{4}{|c|}{\textsc{II. Equation of state}} \\ \hline 
BSK24  &  \textbf{0.5323}  &  BSK24  &  \textbf{0.5409} \\
GM1A  &  0.4265  &  GM1A  &  0.4054 \\
SLy4  &  0.0413  &  SLy4  &  0.0537 \\
 \hline \multicolumn{4}{|c|}{\textsc{III. Mass ($\rm M_\odot$)}} \\ \hline 
1.4  &  0.0101  &  1.4  &  0.1226 \\
1.6  &  0.3512  &  1.6  &  \textbf{0.5682} \\
1.8  &  \textbf{0.6387}  &  1.8  &  0.3092 \\
 \hline \multicolumn{4}{|c|}{\textsc{IV. $B_p$ (G)}} \\ \hline 
$1{\times}10^{12}$  &  0.0000  &  $1{\times}10^{12}$  &  0.0157 \\
$3{\times}10^{12}$  &  0.0082  &  $3{\times}10^{12}$  &  0.2136 \\
$5{\times}10^{12}$  &  0.0544  &  $5{\times}10^{12}$  &  0.1813 \\
$7{\times}10^{12}$  &  0.0725  &  $7{\times}10^{12}$  &  \textbf{0.2806} \\
$1{\times}10^{13}$  &  0.2674  &  $1{\times}10^{13}$  &  0.2015 \\
$2{\times}10^{13}$  &  \textbf{0.4195}  &  $2{\times}10^{13}$  &  0.1072 \\
$3{\times}10^{13}$  &  0.1476  &  $3{\times}10^{13}$  &  0.0001 \\
$5{\times}10^{13}$  &  0.0304  &  $5{\times}10^{13}$  &  0.0000 \\
$7{\times}10^{13}$  &  0.0000  &  $7{\times}10^{13}$  &  0.0001 \\

\hline
    \end{tabular}
    \caption{For the two sources, \psrb and \psra, the five most probable models in the 4D space, sorted by their classification probability (panel \emph{I}), as well as the marginalised probabilities of the considered EoSs, and values for the mass and magnetic field (panels \emph{II}--\emph{IV}). The highest marginal probabilities are denoted with bold typeface.}
    \label{tab:classprob4D}
     \endgroup
\end{table*}

\newpage



\renewcommand\thefigure{\arabic{figure}} 
\setcounter{figure}{0}
\renewcommand{\figurename}{Supplementary Data Figure}
\renewcommand{\tablename}{Supplementary Data Table}

\begin{figure*}[htp!]
	\includegraphics[width=0.48\textwidth]{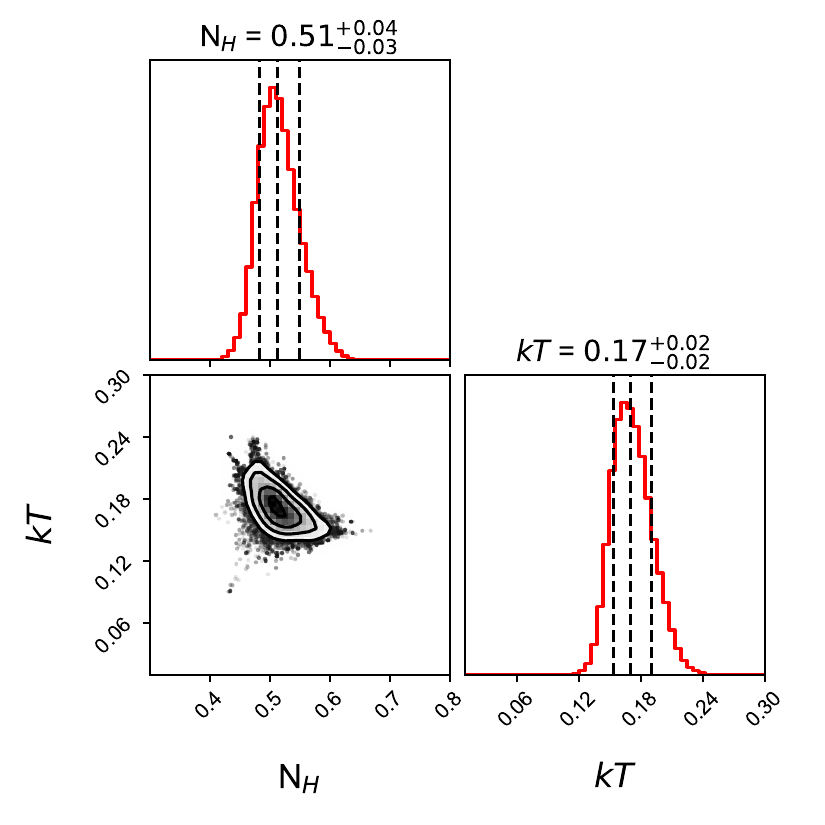}
    \includegraphics[width=0.48\textwidth]{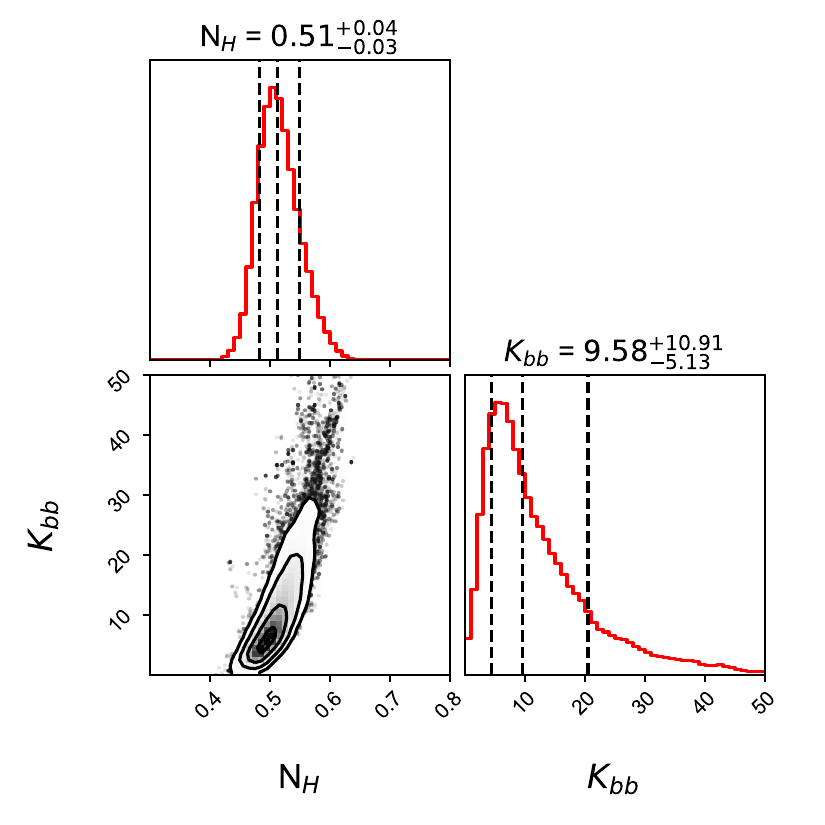}
    \caption{Posterior probability distributions for $N_H$, $kT_{\rm bb}$ and the \textsc{bbodyrad} normalization ${\rm K}_{\rm bb}$ for \psra. Contours represent the 1$\sigma$ , 2$\sigma$ and 3$\sigma$ confidence levels. Marginal posterior distributions are shown as histograms with the median and 1 $\sigma$ intervals of confidence highlighted as dashed lines.}
    \label{fig:mcmc}
\end{figure*}

\newpage

\begin{figure}[htp!]
	\includegraphics[width=1.0\columnwidth]{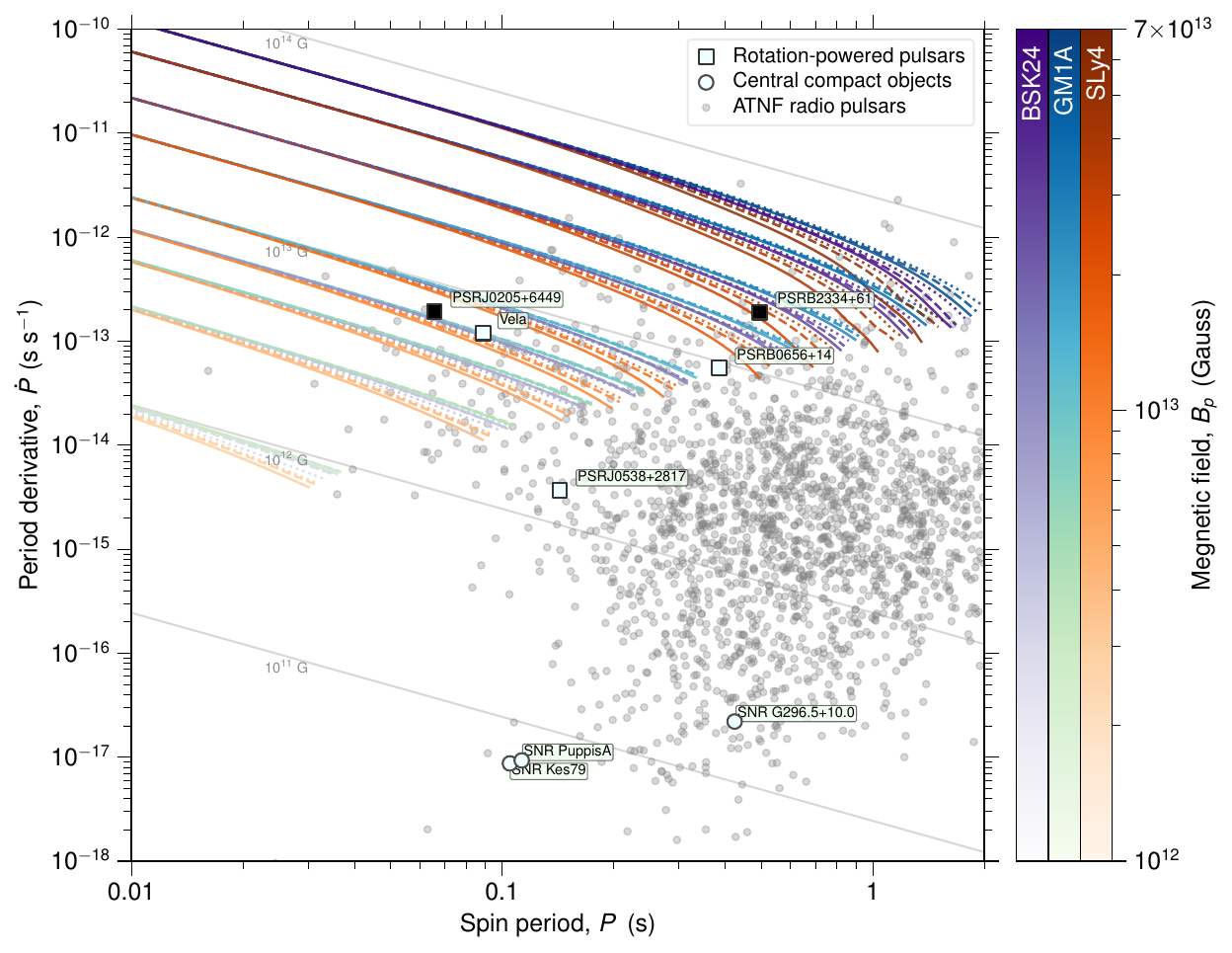}
    \caption{Comparison between observational data and theoretical cooling curves in the $P$-$\dot{P}$ diagram. The curves and data points are stylised following the same prescriptions as in Fig.1. The sources used in this work are highlighted in black.
}
    \label{fig:p_pdot}
\end{figure}

\newpage

\begin{figure}[htp!]
    \includegraphics[width=\columnwidth]{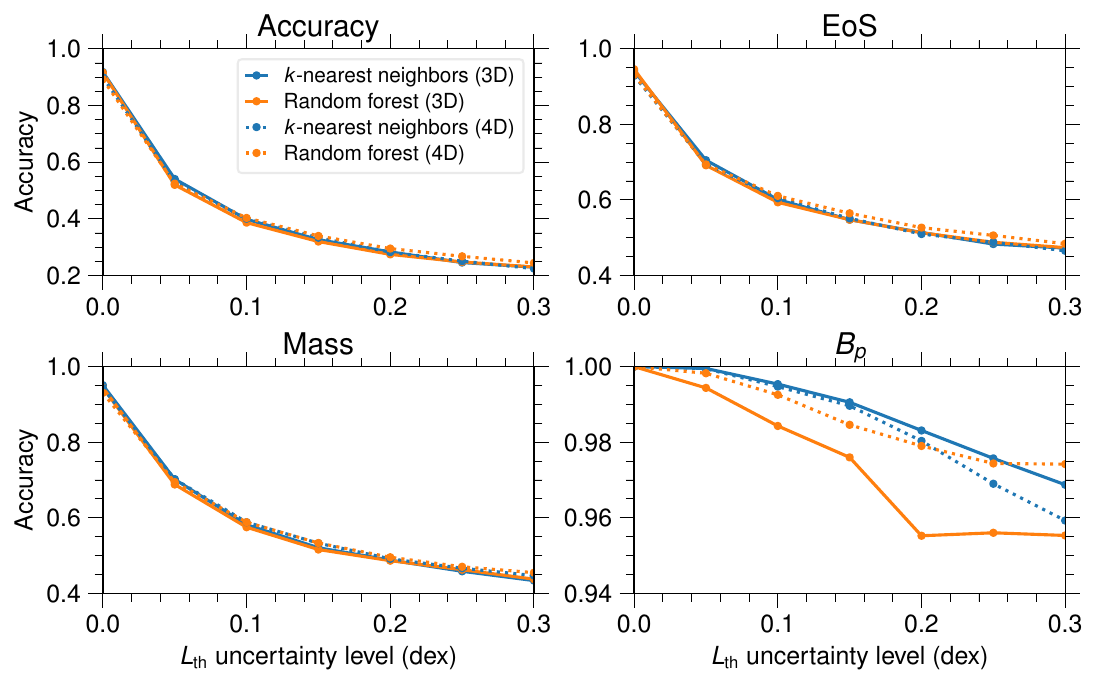}
    \caption{The accuracy score (top left panel; \emph{Accuracy}) of the best-performing 3D (solid lines) and 4D (dotted lines) classifiers, $k$-nearest neighbours (blue) and random forest (orange), as a function of the uncertainty scale in the $L_{\rm th}$. In the rest of the panels we show the marginalised accuracy for the EoS, mass and $B_p$ (see titles).}
    \label{fig:accuracy}
\end{figure}

\newpage

\begin{figure}[htp!]
    \includegraphics[width=\columnwidth]{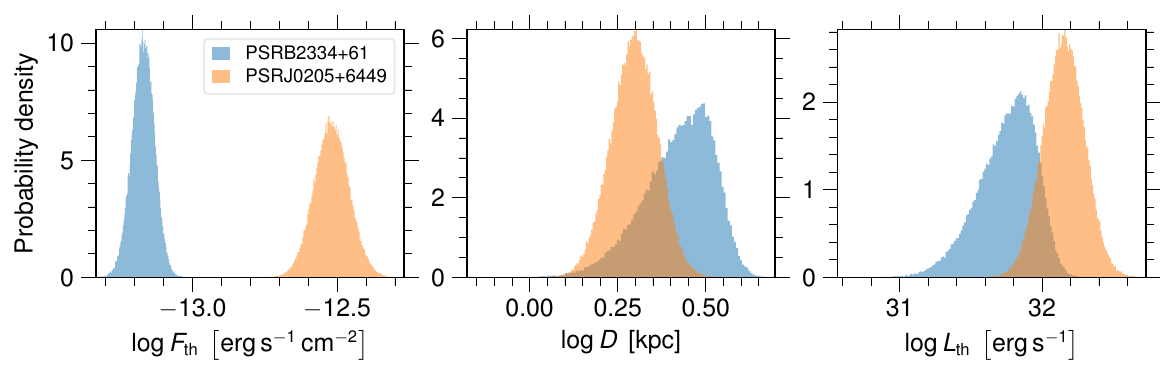}
    \caption{Histograms of samples from the distributions of the flux ($F_{\rm th}$; left), distance (middle) and luminosity ($L_{\rm th}$; right) in logarithmic space, for both sources (PSRB2334+61 with blue, and PSRJ0205+6449 with orange). The shapes of the error distributions illustrate the asymmetry in the independent variables (flux and distance) and the Monte Carlo-propagated luminosity.}
    \label{fig:lx_samples}
\end{figure}

\newpage

\begin{figure*}
    \centering
    \includegraphics[width=\columnwidth]{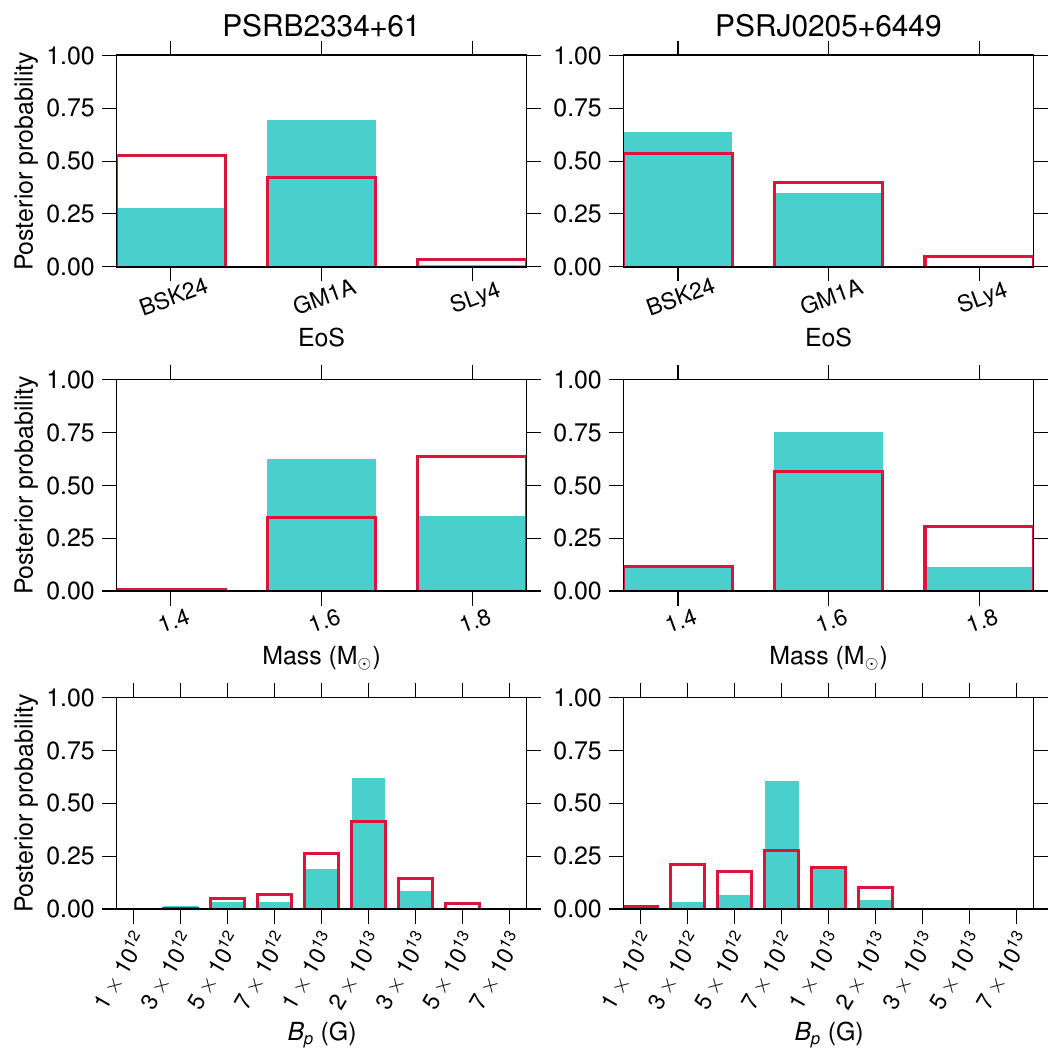}
    \caption{The marginalised probabilities for the EoS, mass and $B_p$ of the two sources, \psrb and \psra, using the 3D (cyan filled bars) and 4D classifiers (red rectangles).}
    \label{fig:posteriors}
\end{figure*}

\newpage


\begin{table*}
\caption{Luminosities \label{tab:luminosity}}
\begin{center}
\begin{tabular}{ l c c c }
\hline 
\hline
Luminosity &  \psra & \psrb & CXOU J0852 \\
\hline
$L_{\rm obs}$ ($10^{32}\rm\,erg\,s^{-1}$) 
    & 1.44$^{+0.6}_{-0.5}$ & 0.78$^{+0.30}_{-0.10}$ & 2.0$\pm$0.6 \\
$L_{\rm stat}$ ($10^{32}\rm\,erg\,s^{-1}$) 
    & $1.42^{+1.02}_{-0.61}$\textsuperscript{\dag}
    & $0.60^{+0.49}_{-0.36}$\textsuperscript{\dag}
    & - \\
$L_{\rm cool}$ ($10^{32}\rm\,erg\,s^{-1}$) & $<1\times10^{-4}$ & $<$0.4 & $<$1.3 \\
\hline
\hline
\end{tabular}
\end{center}
Quoted errors reflect 90\% confidence levels. See uncertainty propagation analysis in Methods for the $L_{\rm stat}$ of two of the sources (\textsuperscript{\dag}).
\end{table*}

\newpage


\begin{table*}
    \centering
    \begingroup
    \footnotesize
    \setlength{\tabcolsep}{8pt} 
    \renewcommand{\arraystretch}{0.7} 
    \begin{tabular}{|l|c||l|c|}
    \hline

\multicolumn{2}{|c||}{PSRB2334+61}  &  \multicolumn{2}{c|}{PSRJ0205+6449} \\
 \hline \multicolumn{4}{|c|}{\textsc{I. Five best models and their classification probabilities}} \\ \hline 
GM1A, $1.6\,M_\odot$, $2{\times}10^{13}$G  &  0.4534  &  BSK24, $1.6\,M_\odot$, $7{\times}10^{12}$G  &  0.3995 \\
BSK24, $1.8\,M_\odot$, $1{\times}10^{13}$G  &  0.1070  &  GM1A, $1.6\,M_\odot$, $7{\times}10^{12}$G  &  0.1064 \\
GM1A, $1.8\,M_\odot$, $2{\times}10^{13}$G  &  0.0990  &  BSK24, $1.6\,M_\odot$, $1{\times}10^{13}$G  &  0.0881 \\
BSK24, $1.6\,M_\odot$, $2{\times}10^{13}$G  &  0.0571  &  GM1A, $1.4\,M_\odot$, $7{\times}10^{12}$G  &  0.0688 \\
GM1A, $1.6\,M_\odot$, $3{\times}10^{13}$G  &  0.0558  &  GM1A, $1.6\,M_\odot$, $1{\times}10^{13}$G  &  0.0546 \\
 \hline \multicolumn{4}{|c|}{\textsc{II. Equation of state}} \\ \hline 
BSK24  &  0.2855  &  BSK24  &  \textbf{0.6426} \\
GM1A  &  \textbf{0.6986}  &  GM1A  &  0.3519 \\
SLy4  &  0.0159  &  SLy4  &  0.0055 \\
 \hline \multicolumn{4}{|c|}{\textsc{III. Mass ($\rm M_\odot$)}} \\ \hline 
1.4  &  0.0093  &  1.4  &  0.1259 \\
1.6  &  \textbf{0.6307}  &  1.6  &  \textbf{0.7577} \\
1.8  &  0.3600  &  1.8  &  0.1165 \\
 \hline \multicolumn{4}{|c|}{\textsc{IV. $B_p$ (G)}} \\ \hline 
$1{\times}10^{12}$  &  0.0002  &  $1{\times}10^{12}$  &  0.0118 \\
$3{\times}10^{12}$  &  0.0171  &  $3{\times}10^{12}$  &  0.0363 \\
$5{\times}10^{12}$  &  0.0390  &  $5{\times}10^{12}$  &  0.0710 \\
$7{\times}10^{12}$  &  0.0400  &  $7{\times}10^{12}$  &  \textbf{0.6111} \\
$1{\times}10^{13}$  &  0.1923  &  $1{\times}10^{13}$  &  0.2069 \\
$2{\times}10^{13}$  &  \textbf{0.6213}  &  $2{\times}10^{13}$  &  0.0484 \\
$3{\times}10^{13}$  &  0.0900  &  $3{\times}10^{13}$  &  0.0102 \\
$5{\times}10^{13}$  &  0.0000  &  $5{\times}10^{13}$  &  0.0041 \\
$7{\times}10^{13}$  &  0.0000  &  $7{\times}10^{13}$  &  0.0001 \\

    \hline
    \end{tabular}
    \caption{For the two sources, \psrb and \psra, the five most probable models in the 3D space, sorted by their classification probability (panel \emph{I}), as well as the marginalised probabilities of the considered EoSs, and values for the mass and magnetic field (panels \emph{II}--\emph{IV}). The highest marginal probabilities are denoted with bold typeface.}
    \label{tab:classprob}
    \endgroup
\end{table*}

\newpage

 \end{document}